\documentclass[twocolumn,aps,pra,amsmath,superscriptaddress,notitlepage,longbibliography]{revtex4-1}
\usepackage{amssymb}
\usepackage{mathrsfs}
\usepackage{graphicx}
\usepackage{float}
\usepackage[caption=false]{subfig}
\usepackage{epstopdf}

\usepackage[normalem]{ulem}
\usepackage{verbatim}
\usepackage{xcolor}
\usepackage{bm}

\usepackage[utf8x]{inputenc}
\usepackage[colorlinks,citecolor=blue]{hyperref}

\begin{document}

\title{Nonlinear Hall effects with an exceptional ring}
\author{Fang Qin}
\email{qinfang@just.edu.cn}
\affiliation{School of Science, Jiangsu University of Science and Technology, Zhenjiang 212100, China}
\author{Ruizhe Shen}
\email{ruizhe20@u.nus.edu}
\affiliation{Department of Physics, National University of Singapore, Singapore 117551, Singapore}
\author{Ching Hua Lee}
\email{phylch@nus.edu.sg}
\affiliation{Department of Physics, National University of Singapore, Singapore 117551, Singapore}
\date{\today}
\begin{abstract}
In non-Hermitian band structures, exceptional points generically form gapless lines or loops that give rise to extensively many defective eigenstates. In this work, we investigate how they nontrivially contribute to higher-order nonlinear responses by introducing unique singularities in the Berry curvature dipole (BCD) or Berry connection polarizability (BCP). Using a tilted two-dimensional dissipative Dirac model ansatz that harbors an exceptional ring, broken inversion symmetry is shown to give rise to extrinsic (BCD) and intrinsic (BCP) nonlinear Hall behaviors unique to systems with extensive exceptional singularities. In particular, when the non-Hermiticity is increased while keeping the ring radius fixed, the BCD response exhibits a power-law increase, while the BCP response correspondingly decreases. Our work sheds light on how non-Hermiticity can qualitatively control the extent and nature of higher harmonic generation in solids.
\end{abstract}
\maketitle

\section{Introduction}\label{section1}

Beyond the well-known Kubo formula paradigm that connects band topology to linear response properties~\cite{kubo1957statistical_1,kubo1957statistical_2,thouless1982quantized,shen2017topological,sticlet2022kubo,qin2020theory,qin2022phase,qin2023light,qin2024kinked,qin2024light,goerbig2009quantum}, the topology and geometry of band crossings also intimately dictates the nonlinear response, particularly in terms of technologically relevant higher harmonic generation properties~\cite{morimoto2016topological,lee2015negative,lee2020enhanced,tai2021anisotropic,mrudul2021high,bharti2022high,bharti2023massless,lv2021high,ghimire2019high,nourbakhsh2021high}.

A key nonlinear response phenomenon is the extrinsic nonlinear Hall effect, which arises due to broken inversion symmetry~\cite{sodemann2015quantum,du2018band,du2019disorder,du2021nonlinear,ortix2021nonlinear,bandyopadhyay2024non}. This unique response is proportional to the Berry curvature dipole (BCD), and has garnered significant experimental interest~\cite{sodemann2015quantum,du2018band,du2019disorder,du2019disorder,du2021nonlinear,ortix2021nonlinear,bandyopadhyay2024non,wang2024nonlinear}, being widely explored in a variety of materials including the two-dimensional (2D) monolayer WTe$_2$~\cite{xu2018electrically}, the bilayer WTe$_2$~\cite{ma2019observation}, the few-layer WTe$_2$~\cite{kang2019nonlinear,xiao2020berry}, the multilayer WTe$_2$~\cite{ye2023control}, the twisted bilayer WTe$_2$~\cite{he2021giant}, the twisted bilayer WSe$_2$~\cite{huang2023giant}, the three-dimensional (3D) Dirac semimetal Cd$_3$As$_2$~\cite{shvetsov2019nonlinear,zhao2023gate}, type-I Weyl semimetal materials~\cite{zhang2018berry,zeng2021nonlinear}, type-II Weyl semimetal materials~\cite{kumar2021room}, the nonmagnetic Weyl--Kondo semimetal~\cite{dzsaber2021giant}, the topological insulator Bi$_2$Se$_3$~\cite{he2021quantum}, the giant Rashba semiconductor BiTeI~\cite{facio2018strongly}, bilayer graphene~\cite{ho2021hall,ma2024skew}, twisted bilayer graphene~\cite{duan2022giant}, twisted double bilayer graphene~\cite{sinha2022berry}, strained twisted bilayer graphene~\cite{pantaleon2021tunable,zhang2022giant}, the strained twisted bilayer WSe$_2$~\cite{hu2022nonlinear}, the strained monolayer WSe$_2$~\cite{qin2021strain}, and the strained 2D monolayer MoS$_2$~\cite{son2019strain}. 
Recently, {\color{blue}there has also been} a surge of interest in the intrinsic nonlinear Hall effect, which alternatively arises from the quantum geometric tensor~\cite{gao2014field,lee2017band,lai2021third,wang2022room,liu2022berry,huang2023intrinsic,gao2023quantum,kaplan2024unification,wang2023quantum,zhuang2023extrinsic,zhuang2024intrinsic,wang2021intrinsic,liu2021intrinsic,fang2024quantum} and has been experimentally probed in thick T$_{d}$-MoTe$_2$ samples~\cite{lai2021third}, Weyl semimetal TaIrTe$_4$~\cite{wang2022room}, and the topological antiferromagnet material MnBi$_2$Te$_4$~\cite{gao2023quantum,kaplan2024unification,wang2023quantum}. Theoretically, the intrinsic nonlinear Hall effect induced by quantum geometry, along with its potential applications in the antiferromagnetic tetragonal CuMnAs~\cite{wang2021intrinsic,liu2021intrinsic} and thin films of the 3D altermagnet RuO$_2$~\cite{fang2024quantum}, has been proposed.

At the same time, non-Hermitian physics has given rise to a new landscape for topological physics~\cite{hatano1996localization,hatano1997vortex,hatano1998non,yao2018edge,yao2018non,kunst2018biorthogonal,yokomizo2019non,lee2019anatomy,deng2019non,kawabata2020non,yokomizo2020non,kawabata2020higher,yi2020non,yokomizo2020topological,xiao2020non,zhang2022universal,yang2020non,zhang2021tidal,yokomizo2021non,bhargava2021non,yokomizo2022non,rafi2022critical,shen2022non,qin2023non,yang2022designing,qin2023universal,li2022non,jiang2023dimensional,tai2023zoology,wan2023observation,wang2024absence,shen2024enhanced,poddubny2024mesoscopic,xue2024topologically,manna2023inner}. The infusion of non-Hermitian topology has greatly enriched the phenomenology of quantum response and transport. Previous studies on non-Hermitian linear response~\cite{qin2024kinked,bouganne2020anomalous,pan2020non,zhao2025universal} have unveiled anomalous coherent oscillations in wave packet dynamics~\cite{xu2017weyl,lee2020unraveling,jiong2021anomalous,lee2020ultrafast,singhal2023measuring,price2012mapping,roy2015wave,oliver2023bloch} induced by the non-Hermitian anomalous Berry connection. As a foundation for our discussion, we introduce the concepts of defective eigenstates, exceptional points, and exceptional rings in non-Hermitian systems. Defective eigenstates emerge at exceptional points, where eigenvalues and eigenvectors coalesce, rendering the Hamiltonian nondiagonalizable and the eigenstates nonorthogonal~\cite{heiss2012the,bergholtz2021exceptional,meng2024exceptional}. Unlike Hermitian degeneracies, such as Dirac or Weyl points, where only eigenvalues become degenerate, exceptional points involve the simultaneous merging of eigenvalues and eigenvectors, leading to a singularity in parameter space~\cite{heiss2012the,bergholtz2021exceptional,meng2024exceptional}. Furthermore, the defectiveness of the exceptional points gives rise to exceptional bound states and results in negative biorthogonal entanglement entropy~\cite{lee2022exceptional,xue2024topologically,zou2024experimental,liu2024non,canos2025exceptional}. An exceptional ring is a continuous loop of exceptional points in momentum space, where eigenvalues and eigenvectors coalesce along a closed trajectory~\cite{heiss2012the,bergholtz2021exceptional,meng2024exceptional}. This phenomenon has been widely studied in photonics~\cite{wang2019exceptional,liu2022experimental,isobe2023a}, cold-atom systems~\cite{xu2017weyl,lei2025topological}, waveguide arrays~\cite{cerjan2019experimental}, and electrical circuits~\cite{cao2020band,sahin2025topolectrical}. In particular, in non-Hermitian systems containing exceptional point singularities, the Berry curvature exhibits singularities near the band crossings, leading to much heightened sensitivity around these singularities and qualitatively affecting both linear and high-order nonlinear responses. In general, the enhancement of nonlinear responses near small band gaps and/or band crossings has been studied in the literature~\cite{ahn2020low,wu2017giant}, and it happens in a variety of Hermitian systems as well. The feature unique to non-Hermitian systems is the possibility of tuning the nonlinear responses with the parameter associated with the non-Hermiticity. As such, in this work, we investigate a tilted two-dimensional massive dissipative Dirac model that represents an ansatz exceptional ring model. Our findings reveal that both the BCD and Berry connection polarizability (BCP) become markedly sensitive to non-Hermiticity near the singularities of an exceptional ring. Particularly, the distinct scaling laws of BCD and BCP in proximity to this exceptional ring lead to significantly different extrinsic and intrinsic nonlinear Hall responses.

The paper is organized as follows: In Section~\ref{section2}, we present the general formalism for both extrinsic and intrinsic nonlinear Hall effects. In Section~\ref{section3}, we introduce the Hamiltonian for an exceptional ring and provide a detailed analysis of its energy band structure. In Section~\ref{section4}, we investigate the quantum geometric tensor, Berry curvature, quantum metric, and the extrinsic nonlinear Hall response described by the BCD. Moreover, we study the nonzero component intrinsic nonlinear Hall response described by the BCP. In Section~\ref{section5}, we discuss the viable mechanisms to engineer non-Hermiticity in solid-state systems. Finally, we provide a conclusion section, i.e., Section~\ref{section6}.

\section{General formalism for nonlinear Hall effects}\label{section2}

In this section, we introduce the foundational framework for understanding both extrinsic and intrinsic nonlinear Hall effects, with a focus on how such quantum geometric properties are modified in non-Hermitian systems.

\subsection{Quantum geometric tensor}\label{section2.1}

The quantum geometric tensor, which combines information about the Berry curvature and quantum metric, plays a crucial role in the topological characterization of responses. Unlike the conventional Hermitian setup, the biorthogonal quantum geometric tensor in a non-Hermitian model is defined using the left and right eigenvectors $|\psi_{R}^{(n)}\rangle$ and $|\psi_{L}^{(n)}\rangle$, of the $n$th energy band \cite{brody2013biorthogonal,lee2022exceptional,lee2020unraveling,liang2013topological,zou2024detecting,yao2018edge,yao2018non,kunst2018biorthogonal,yokomizo2019non,lee2019anatomy,deng2019non,kawabata2020non,yokomizo2020non,kawabata2020higher,yi2020non,yokomizo2020topological,xiao2020non,zhang2022universal,yang2020non,zhang2021tidal,yokomizo2021non,bhargava2021non,yokomizo2022non,rafi2022critical,shen2022non,qin2023non,yang2022designing,qin2023universal,li2022non,jiang2023dimensional,tai2023zoology,wan2023observation,wang2024absence,shen2024enhanced,poddubny2024mesoscopic,xue2024topologically}. This tensor can be defined as a matrix in the $r$-dimensional subspace spanned by $r$ selected occupied bands~\cite{provost1980riemannian,resta2011insulating,ma2010abelian,lee2017band,ezawa2024analytic,ghosh2024probing,torma2022superconductivity,peotta2025quantum,torma2023essay,hetenyi2023fluctuations,gong2024nonlinear,hu2024generalized,hu2025quantum,behrends2025quantum}:
\begin{eqnarray}
{\cal Q}^{(nm)}_{ab}
&\!=\!&\langle\partial_{k_a}\psi_{L}^{(n)}|(\mathbb{I}\!-\!\hat{P})|\partial_{k_b}\psi_{R}^{(m)}\rangle \nonumber\\
&\!=\!&{\cal G}^{(nm)}_{ab}\!-\!\frac{i}{2}\Omega_{ab}^{(nm)},\label{eq:QGT}
\end{eqnarray} where $a,b\in\{x,y,z\}$ are for  spatial coordinates, $\mathbb{I}$ is the identity matrix, $\hat{P}\!=\!\sum_{n\rq{}=1}^{r}|\psi_{R}^{(n\rq{})}\rangle\langle\psi_{L}^{(n\rq{})}|$ is the biorthogonal projection operator~\cite{onishi2024fundamental} so that ${\cal Q}_{ab}$ is a $r\times r$ matrix. Here, $n\rq{}$ denotes the index of the selected occupied bands and $r$ denotes the total number of the selected occupied bands. Here, ${\cal G}^{(nm)}_{ab}$ and $\Omega_{ab}^{(nm)}$ are the biorthogonal quantum metric and Berry curvature respectively, which are given by~\cite{ma2010abelian,ezawa2024analytic,onishi2024fundamental,matsuura2010momentum,von2021measurement,yang2024percolation,chen2024quantum,ezawa2024intrinsic,liu2024quantum}
\begin{eqnarray}
{\cal G}^{(nm)}_{ab}&\!\equiv\!&\frac{1}{2}{\rm Re}\left[{\cal Q}^{(nm)}_{ab} \!+\! {\cal Q}^{(mn)}_{ba} \right] \nonumber\\
&\!=\!&\frac{1}{2}{\rm Re}\left[\langle\partial_{k_a}\psi_{L}^{(n)}|\partial_{k_b}\psi_{R}^{(m)}\rangle \!+\! \langle\partial_{k_b}\psi_{L}^{(m)}|\partial_{k_a}\psi_{R}^{(n)}\rangle \right] \nonumber\\
&&\!- \frac{1}{2}\sum_{n\rq{}=1}^{r}{\rm Re}\left[\! \langle\partial_{k_a}\psi_{L}^{(n)}|\psi_{R}^{(n\rq{})}\rangle\langle\psi_{L}^{(n\rq{})}|\partial_{k_b}\psi_{R}^{(m)}\rangle \right.\nonumber\\
&&\left.\!+ \langle\partial_{k_b}\psi_{L}^{(m)}|\psi_{R}^{(n\rq{})}\rangle\langle\psi_{L}^{(n\rq{})}|\partial_{k_a}\psi_{R}^{(n)}\rangle \!\right]\!, \label{eq:QM1}
\end{eqnarray}
\begin{eqnarray}
\Omega_{ab}^{(nm)}&\!\equiv\!&{\rm Re}\left\{i\left[{\cal Q}^{(nm)}_{ab} \!-\! {\cal Q}^{(mn)}_{ba} \right]\right\} \nonumber\\
&\!=\!&{\rm Re}\left\{i\left[\! \langle\partial_{k_a}\psi_{L}^{(n)}|\partial_{k_b}\psi_{R}^{(m)}\rangle \!-\! \langle\partial_{k_b}\psi_{L}^{(m)}|\partial_{k_a}\psi_{R}^{(n)}\rangle \!\right]\right\} \nonumber\\
&&\!- \sum_{n\rq{}=1}^{r}{\rm Re}\left\{i\left[\! \langle\partial_{k_a}\psi_{L}^{(n)}|\psi_{R}^{(n\rq{})}\rangle\langle\psi_{L}^{(n\rq{})}|\partial_{k_b}\psi_{R}^{(m)}\rangle \right.\right.\nonumber\\
&&\left.\left.\!- \langle\partial_{k_b}\psi_{L}^{(m)}|\psi_{R}^{(n\rq{})}\rangle\langle\psi_{L}^{(n\rq{})}|\partial_{k_a}\psi_{R}^{(n)}\rangle \!\right]\!\right\}\!,\label{eq:BC1}
\end{eqnarray} where we have the relations $\Omega_{ba}^{(mn)}\!=\!-\Omega_{ab}^{(nm)}$. Here, we have $\Omega_{c}^{(nm)}\!=\!\varepsilon^{abc}\Omega_{ab}^{(nm)}$, where $\varepsilon^{abc}$ is the Levi-Civita antisymmetric tensor, and $a$, $b$, and $c$ represent the spatial coordinates $x$, $y$, and $z$.

In Section~\ref{section2.1}, Eqs.~\eqref{eq:QGT}-\eqref{eq:BC1} are presented in their general tensor form for multiband systems. However, in Section~\ref{section2.2} and Section~\ref{section2.3}, when calculating the linear and nonlinear response quantities, we specifically focus on the diagonal terms and selected occupied bands in Eqs. \eqref{eq:QGT}-\eqref{eq:BC1}. In other words, our analysis in Section~\ref{section2.2} and Section~\ref{section2.3} considers only the case where $n\!=\!m$ in these equations.

\subsection{Extrinsic nonlinear Hall effect}\label{section2.2}

We now examine the extrinsic nonlinear Hall effect, where the direct response to an external force manifests in the velocity response of affected particles. Here, we consider a simple scenario involving a time-driven oscillating force 
\begin{equation}
{\bf F}(t)=F_{a}^{\omega}(t){\bf e}_{a} = {\rm Re}(\mathcal{F}_{a}e^{i\omega t}){\bf e}_{a}
\label{eq:force}
\end{equation}
along the $a$ direction. $\bold F(t)$ has the unit of force, and it enters the nonequilibrium distribution function $f$ through the Boltzmann equation within the relaxation time approximation, which is a semiclassical approximation. 
This force can induce an additional transverse velocity along the $b$ direction in the particle clouds~\cite{sodemann2015quantum,du2018band,du2019disorder,lee2018floquet,qin2024light}:
\begin{equation}
\bar{v}_{b}^{}(\mathcal{F}_{a})\approx{\rm Re}\left(\bar{v}_{b}^{0}+\bar{v}_{b}^{\omega}e^{i\omega t} + \bar{v}_{b}^{2\omega}e^{i2\omega t}\right).\label{eq:v}
\end{equation} 
We first discuss the linear contribution $\bar{v}_{b}^{\omega}\!=\!\chi_{ba}^{(1)}\mathcal{F}_{a}$, where the coefficient $\chi_{ba}^{(1)}\!\approx\!\frac{{\cal S}}{2\pi\hbar}C_{ba}\!=\!\frac{{\cal S}}{2\pi\hbar}\varepsilon_{bac}C_{c}$ with~\cite{qin2024light}
\begin{equation}
C_{c}\!=\!\varepsilon^{bac}C_{ba}\!=\!-C_{ba}\!=\!-2\pi\sum_{n}\int\frac{d^{2}{\bf k}}{(2\pi)^2}\Omega_{c}^{(n)}f_{0}^{(n)}.\label{eq:Cc_0}
\end{equation} 
Here $\mathcal{F}_{a}$ is the magnitude of force along the $a$ direction, ${\cal S}$ is the area of the system, $\hbar$ is the reduced Planck\rq{}s constant.
In Eq.~\eqref{eq:Cc_0}, $\Omega_{c}^{(n)}$ is the $c$th component of the Berry curvature ${\bf\Omega}^{(n)}$ computed for the $n$th eigenstate, and $f_{0}^{(n)}\!=\!f_{0}^{(n)}(\epsilon_{\bf k}^{(n)}\!-\!E_{F})$ is the equilibrium Fermi-Dirac distribution function of the $n$th eigenenergy band $\epsilon_{\bf k}^{(n)}$ and Fermi energy $E_{F}$.
Specifically, under the oscillating force ${\mathbf{F}}\!=\!F_{a}^{\omega}(t){\bf e}_{a}$ with $x\!=\!a$ [Eq.~\eqref{eq:force}]~\cite{aidelsburger2015measuring,mancini2015observation,goldman2016topological,cooper2019topological}, we obtain the linear velocity contribution in the $y$ direction: 
\begin{equation}
\bar{v}_{y}^{\omega}\!\approx\!\frac{{\cal S}}{2\pi\hbar}\mathcal{F}_{x}C_{z}.\label{eq:vy_L}
\end{equation} 

Next, we focus on the nonlinear response, which arises from the following two contributions: $\bar{v}_{b}^{0}\!=\!\chi_{baa}^{(0)}\mathcal{F}_{a}^{2}$ and $\bar{v}_{b}^{2\omega}\!=\!\chi_{baa}^{(2)}\mathcal{F}_{a}^{2}$. The coefficients are given by $\chi_{baa}^{(0)}\approx
\chi_{baa}^{(2)}\!\approx\! \frac{\tau{\cal S}D_{baa}}{2\hbar^2(1+i\omega\tau)}$~\cite{sodemann2015quantum,du2018band,du2019disorder}, where $\tau$ is the relaxation time~\cite{sodemann2015quantum,du2018band,du2019disorder}. At the core of the extrinsic nonlinear response is the BCD $D_{baa}$, as given by~\cite{sodemann2015quantum,du2018band,du2019disorder,zhu2024nonlinear,chen2024nonlinear}: 
\begin{equation}
D_{baa}\!=\!\sum_{n}\int\frac{d^{2}{\bf k}}{(2\pi)^2}\Omega_{ba}^{(n)}(\partial_{k_a}\epsilon_{\bf k}^{(n)})\frac{\partial f_{0}^{(n)}}{\partial\epsilon_{\bf k}^{(n)}}.\label{eq:BCD_0}
\end{equation} 
The derivative of the equilibrium distribution function $\partial f_{0}^{(n)}/\partial\epsilon_{\bf k}^{(n)}$ in the above integral for the BCD indicates that the extrinsic nonlinear Hall response is predominantly governed by the states near the Fermi surface \cite{du2018band,du2019disorder,du2021quantum,du2021nonlinear,xiao2019theory,zhu2024nonlinear,chen2024nonlinear}.
Setting $a\!=\!x$ in Eq.~\eqref{eq:force}, the extrinsic nonlinear response manifesting as the $y$-direction nonlinear velocity can be obtained by solving the Boltzmann equation under the relaxation time approximation~\cite{sodemann2015quantum,du2018band,du2019disorder}, 
\begin{equation}
\bar{v}_{yxx}^{\rm BCD}\!\approx\!\bar{v}_{y}^{0}\!\approx\!\bar{v}_{y}^{2\omega}\!\approx\!\frac{\tau{\cal S}\mathcal{F}_{x}^{2}}{2\hbar^{2}}D_{yxx},\label{eq:vy_NL}
\end{equation} 
where $\bar{v}_{yxx}^{\rm BCD}$ is the extrinsic nonlinear velocity, and we have used $\omega\tau\to0$. Furthermore, with the help of Eq.~\eqref{eq:BCD_0}, we obtain
\begin{eqnarray}
D_{yxx}
&\!=\!&\sum_{n}\int\frac{d^{2}{\bf k}}{(2\pi)^2}\!\left[\!\Omega_{yx}^{(n)}\left(\partial_{k_x}\epsilon_{\bf k}^{(n)}\right)\frac{\partial f_{0}^{(n)}}{\partial\epsilon_{\bf k}^{(n)}}\right]\!\! \nonumber\\
&\!=\!&\!-\!\sum_{n}\int\frac{dk_y}{2\pi}\!\left[f_{0}^{(n)}\Omega_{z}^{(n)}\right]\bigg|_{k_x=-\pi}^{k_x=\pi}\nonumber\\
&&\!+\sum_{n}\int\frac{d^{2}{\bf k}}{(2\pi)^2}\!\left[f_{0}^{(n)}\frac{\partial\Omega_{z}^{(n)}}{\partial k_x}\right]\!,\label{eq:BCD}
\end{eqnarray} where we used $\Omega_{yx}^{(n)}\!=\!-\Omega_{xy}^{(n)}\!=\!-\Omega_{z}^{(n)}$. Note that the inversion-symmetry breaking is essential for the finite Berry curvature and nonvanishing BCD \cite{du2018band,du2019disorder,du2021quantum,du2021nonlinear,xiao2019theory,chen2024nonlinear}.

\subsection{Intrinsic nonlinear Hall effect}\label{section2.3}

The second-order velocity tensor (nonlinear velocity tensor) is defined as the contribution to the velocity response that is quadratic to the applied force field ${\bf F}$ as $\bar{v}_{abc}\!\propto\!\mathcal{F}_{b}\mathcal{F}_{c}$~\cite{gao2014field}, where $\bar{v}_{abc}$ can be separated into an Ohmic-type nonlinear response and a Hall-type nonlinear response~\cite{tsirkin2022separation}. The Ohmic-type nonlinear response includes a second-order Drude conductivity, which is quadratically dependent on the relaxation time~\cite{holder2020consequences,watanabe2020nonlinear,watanabe2021chiral,vzelezny2021unidirectional}. The Hall-type nonlinear response is independent of the relaxation time and is therefore called the intrinsic nonlinear Hall effect. 

This intrinsic nonlinear response can be described by the BCP~\cite{gao2014field,wang2021intrinsic,liu2021intrinsic,lai2021third,liu2022berry}, as given by the intrinsic nonlinear velocity $\bar{v}_{abc}^{\rm BCP}$~\cite{gao2014field,liu2021intrinsic,wang2021intrinsic,liu2022berry,zhuang2023extrinsic,zhuang2024intrinsic,chen2024nonlinear}
\begin{eqnarray}
\bar{v}_{abc}^{\rm BCP}
&\!=\!&\!\!\sum_{n,n\rq{}}^{\epsilon_{\bf k}^{(n)}\neq \epsilon_{\bf k}^{(n\rq{})}}\!\Gamma_{bc}\!\int\!\!\frac{d^{2}\mathbf{k}}{(2\pi)^{2}}\!\!\left[\frac{v_{a}^{(n)}{\cal G}_{bc}^{(n)}\!-\!v_{b}^{(n)}{\cal G}_{ac}^{(n)}}{\epsilon_{\bf k}^{(n)}\!-\!\epsilon_{\bf k}^{(n\rq{})}}\right]\!\!\frac{\partial f_{0}^{(n)}}{\partial\epsilon_{\bf k}^{(n)}} \nonumber\\
&\!=\!&\sum_{n,n\rq{}}^{\epsilon_{\bf k}^{(n)}\neq \epsilon_{\bf k}^{(n\rq{})}}\!\Gamma_{bc}\!\int\frac{dk_b}{2\pi}\!\left[\frac{f_{0}^{(n)}{\cal G}_{bc}^{(n)}}{\epsilon_{\bf k}^{(n)}\!-\!\epsilon_{\bf k}^{(n\rq{})}}\right]\bigg|_{k_a=-\pi}^{k_a=\pi}  \nonumber\\
&&\!+\! \sum_{n,n\rq{}}^{\epsilon_{\bf k}^{(n)}\neq \epsilon_{\bf k}^{(n\rq{})}}\!\Gamma_{bc}\!\int\frac{dk_a}{2\pi}\!\left[\frac{-f_{0}^{(n)}{\cal G}_{ac}^{(n)}}{\epsilon_{\bf k}^{(n)}\!-\!\epsilon_{\bf k}^{(n\rq{})}}\right]\bigg|_{k_b=-\pi}^{k_b=\pi} \nonumber\\
&&\!-\! \sum_{n,n\rq{}}^{\epsilon_{\bf k}^{(n)}\neq \epsilon_{\bf k}^{(n\rq{})}}\!\Gamma_{bc}\!\int\!\!\frac{d^{2}\mathbf{k}}{(2\pi)^{2}}f_{0}^{(n)}\!\!\left[\!\frac{\partial_{k_a}{\cal G}_{bc}^{(n)}\!-\!\partial_{k_b}{\cal G}_{ac}^{(n)}}{\epsilon_{\bf k}^{(n)}\!-\!\epsilon_{\bf k}^{(n\rq{})}}\!\right]\!\! \nonumber\\
&&\!+\! \sum_{n,n\rq{}}^{\epsilon_{\bf k}^{(n)}\neq \epsilon_{\bf k}^{(n\rq{})}}\!\Gamma_{bc}\!\int\!\!\frac{d^{2}\mathbf{k}}{(2\pi)^{2}}\!\!\frac{f_{0}^{(n)}}{\left(\epsilon_{\bf k}^{(n)}\!-\!\epsilon_{\bf k}^{(n\rq{})}\right)^{2}} \nonumber\\
&&\times\left[{\cal G}_{bc}^{(n)}\!\partial_{k_a}\!\!\left(\epsilon_{\bf k}^{(n)}\!-\!\epsilon_{\bf k}^{(n\rq{})}\right) \!-\! {\cal G}_{ac}^{(n)}\!\partial_{k_b}\!\!\left(\epsilon_{\bf k}^{(n)}\!-\!\epsilon_{\bf k}^{(n\rq{})}\right)\right], \label{eq:BCP_abc}\nonumber\\
\end{eqnarray} where $\Gamma_{bc}\!=\!2{\cal S}\mathcal{F}_{b}\mathcal{F}_{c}/\hbar$ with $\mathcal{F}_{b/c}$ the force magnitudes along the $b/c$ direction, and $v_{a}^{(n)}\!=\!\partial\epsilon_{\bf k}^{(n)}/\partial k_{a}$ the group velocity of the $n$th band~\cite{zhuang2024intrinsic}. Details on the derivation of Eq.~\eqref{eq:BCP_abc} can be found in Appendix~\ref{Appendix_2}.

\section{Exceptional ring model}\label{section3}

As previously introduced, non-Hermitian systems can exhibit heightened sensitivity near exceptional points~\cite{xu2017weyl,cerjan2019experimental,liu2021higher,zhen2015spawning,kawabata2019classification,yokomizo2020topological,wang2021simulating,ding2016emergence,alvarez2018non,zhou2018observation,xiao2021observation,bergholtz2021exceptional,zou2024detecting,zou2024experimental,lee2022exceptional,wiersig2020review,parto2020non,hu2022knot,meng2022terahertz,ding2022non,mao2024exceptional,meng2024exceptional}. Due to the greater number of non-Hermitian degrees of freedom, exceptional points occur far more generically than usual Hermitian nodal points~\cite{chen2020revealing}, commonly appearing as extended one-dimensional (1D) exceptional lines or loops. Here, we investigate a specific scenario: a system with a ring of exceptional points in its  Brillouin zone, such that its Hamiltonian matrix is not of full rank along a 1D locus. Along this ring, the conduction and valence bands coalesce, and the absence of a complete set of left eigenvectors results in ill-defined Berry curvature with divergent behavior, as later given by Eq.~\eqref{eq:BC}. Prior research on exceptional structures has focused mainly on their band topology~\cite{ding2016emergence,zhou2018observation,kozii2017non,heiss2012the,zhang2021tidal,xiao2021observation,bergholtz2021exceptional,kawabata2019classification,arouca2020unconventional,chang2020entanglement,lee2022exceptional,zou2024experimental,liu2024non}, with limited insights into transport behaviors. This motivates our deeper investigation into the higher-order effects of Berry curvature, specifically the BCD and the BCP.

With this motivation, we introduce a tilted exceptional ring model to investigate the nonlinear exceptional response. Specifically, we consider a two-dimensional tilted massive Dirac model with gain and loss
\begin{equation}
\hat{\cal H}({\bf k})\!=\!d_{0}\sigma_{0} \!+\! \sum_{i=x,y,z}^{}d_{i}\sigma_{i},\label{eq:H}
\end{equation}
where $d_{0}\!=\!t_{0}k_{x}$, $d_{x}\!=\!vk_{y}$, $d_{y}\!=\!\eta vk_{x}$, and $d_{z}\!=\!i\gamma\!+\!M\!-\!\alpha k^{2}$.
Here, $\sigma_{x,y,z}$ are the Pauli matrices and $\sigma_{0}$ denotes the $2\times2$ identity matrix. We set $\eta\!=\!-1$ throughout in our numerical calculations. The parameter $\gamma$ represents the strength of gain or loss~\cite{li2019observation,lapp2019engineering,ren2022chiral,liang2022dynamic,gou2020tunable}. The tilted term $t_{0}k_{x}\sigma_{0}$ breaks the inversion symmetry and breaks the rotational symmetry in the shape of the Fermi surface, playing a pivotal role in triggering the nonlinear Hall effect in our model.

\begin{figure}[htpb]
\centering 
\includegraphics[width=\columnwidth]{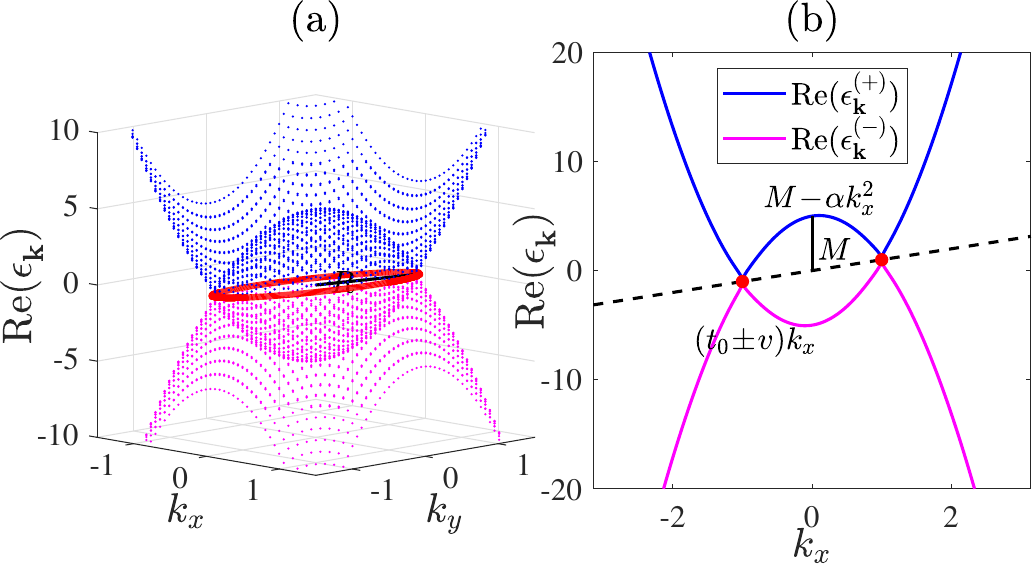}
\caption{(a) Three-dimensional plot of the real energy band structure [Eq.~\eqref{eq:energy}] of our exceptional ring model Eq.~\eqref{eq:H}. (b) The $k_{y}\!=\!0$ slice of the real part of the energy band structure, shown for parameters as $t_{0}\!=\!1$, $\gamma\!=\!v\!=\!1$, $M\!=\!\alpha\!=\!5$ that fixes the ring radius as $R\!=\!\gamma/v\!=\!\sqrt{M/\alpha}\!=\!1$.
}\label{fig:E3D_t02_g1_M5}
\end{figure}

For our non-Hermitian Hamiltonian in Eq.~\eqref{eq:H}, we can solve the eigenenergies for the upper ($+$) and lower ($-$) bands as follows:
\begin{equation}
\epsilon_{\mathbf{k}}^{(\pm)}\!=\!t_{0}k_{x}\!\pm\!\sqrt{(i\gamma\!+\!M\!-\!\alpha k^{2})^{2}\!+\!v^{2}k^{2}},\label{eq:energy}
\end{equation}
where $k^{2}\!=\!k^{2}_{x}\!+\!k^{2}_{y}$. In this model, the bands touch along an exceptional ring, depicted in red as shown in Fig.~\ref{fig:E3D_t02_g1_M5}(a). The corresponding $k_{y}\!=\!0$ slice of the energy band structure is also plotted in Fig.~\ref{fig:E3D_t02_g1_M5}(b) to show the gapless crossings more clearly. The exceptional ring in our model, where the degeneracy of eigenvalue vanishes \cite{zhen2015spawning,xu2017weyl,cerjan2019experimental,kawabata2019classification,zhang2021tidal,liu2021higher}: $(i\gamma\!+\!M\!-\!\alpha k^{2})^{2}\!+\!v^{2}k^{2}\!=\!0$, is fixed at
\begin{equation}
k^{2}\!=\!R^{2}\!=\!M/\alpha\!=\!\gamma^{2}/v^{2}, \label{eq:ER_conditoins}
\end{equation} where $R$ is the radius of the exceptional ring in momentum space, and can be written as $R\!=\!\sqrt{M/\alpha}\!=\!\gamma/v$. 
Physically, $M$ represents the Dirac mass, with $2M$ describing the energy gap at $(k_x,k_y)\!=\!(0,0)$. In particular, when $M\!=\!0$, a half-quantized Hall conductance emerges at the center of the Dirac cone~\cite{qin2023light}. The parameter $\alpha$ accounts for the modified quadratic momentum-dependent Dirac mass, expressed as $M\!-\!\alpha k^{2}$, while the parameter $t_{0}\!\pm\!v$, characterizes the slope of the exceptional point's linear energy spectrum, where $v$ is the strength of the Rashba spin-orbit coupling.

Further properties of this exceptional ring are described in the appendices. Specifically, as detailed in Appendix~\ref{Appendix_3}, we demonstrate that the Hamiltonian~\eqref{eq:H} retains inversion symmetry at $t_{0}\!=\!0$, which is however broken under $t_{0}\!\neq\!0$. Moreover, in Appendix~\ref{Appendix_4}, we establish that the Hamiltonian~\eqref{eq:H} does not exhibit time-reversal symmetry, irrespective of the value of $t_{0}$. To explore the properties of the model near the exceptional ring, we provide analytical expansions of both the Hamiltonian and the eigenenergies near these points in Appendix~\ref{Appendix_5}. In the following section, we proceed to explore the nonlinear responses with such an exceptional ring.

\begin{figure*}[htpb]
\centering
\includegraphics[width=.8\textwidth]{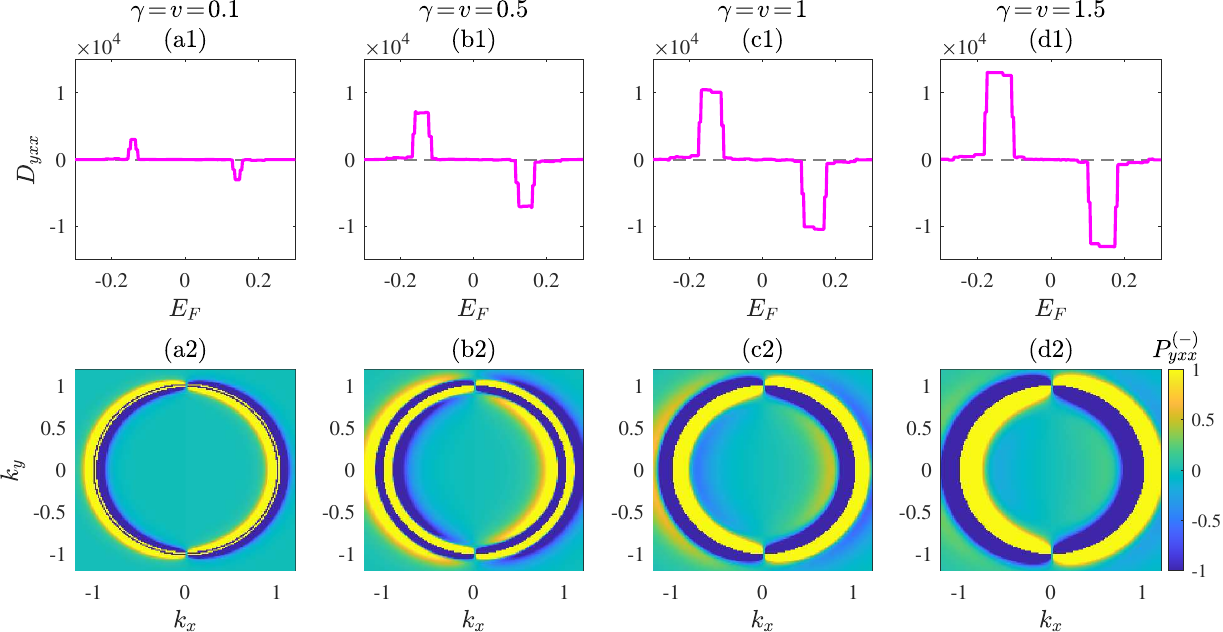}
\caption{\textbf{Extrinsic nonlinear response of our exceptional ring with fixed radius $R\!=\!\gamma/v\!=\!\sqrt{M/\alpha}\!=\!1$, with $M\!=\!\alpha\!=\!5$.} (a1-d1): The Berry curvature dipole (BCD) [Eq.~\eqref{eq:BCD}], showing strong peaks that increase with the non-Hermiticity $\gamma$. (a2)-(d2): The BCD density profile $P_{yxx}^{(-)}({\bf k})$ [Eq.~\eqref{eq:BC_kx}] in the $k_{x}$--$k_{y}$ plane. The band structure tilt is $t_{0}\!=\!0.2$.}
\label{fig:BCD_Pxz_t02_g_v_together_M5}
\end{figure*}

\subsection{Model Berry curvature and quantum metric}\label{section4.1}
In our work, we investigate both the extrinsic and intrinsic nonlinear Hall effects. Here, we first focus on the extrinsic nonlinear response. This arises from the biorthogonal Berry curvature which, with the help of Eq.~\eqref{eq:BC1}, can be expressed for our non-Hermitian model \eqref{eq:H} as follows~\cite{kubo1957statistical_1,kubo1957statistical_2,thouless1982quantized,shen2017topological,sticlet2022kubo,qin2023light,qin2024kinked,qin2024light}:
\begin{eqnarray}
\Omega^{(\pm)}_{z}
&\!=\!&\varepsilon^{xyz}\Omega^{(\pm)}_{xy} \nonumber\\
&\!=\!&\pm{\rm Re}\left\{\!\frac{\eta v^2 \left(i\gamma \!+\! M \!+\! \alpha k^2\right)}{2 \left[\left(i\gamma \!+\! M \!-\! \alpha k^2\right)^2 \!+\! v^2 k^2\right]^{3/2}} \!\right\}.\label{eq:BC}
\end{eqnarray} Equation~\eqref{eq:BC} indicates that the Berry curvature diverges at the exceptional ring  where~Eq.~\eqref{eq:ER_conditoins} is satisfied, greatly enhancing the BCD there. Details on the derivation of Eq.~\eqref{eq:BC} can be found in Appendix~\ref{Appendix_6}.

According to Eq.~\eqref{eq:BCP_abc}, the intrinsic nonlinear Hall effect arises from the biorthogonal quantum metric. This quantum metric for our two-band non-Hermitian Dirac model~\eqref{eq:H} can be expressed as~\cite{zhuang2023extrinsic,zhuang2024intrinsic,ezawa2024analytic,onishi2024fundamental,matsuura2010momentum,von2021measurement,chen2024quantum} 
\begin{eqnarray}
{\cal G}^{(\pm)}_{ab}
\!=\!\frac{1}{4}{\rm Re}\left[(\partial_{k_a}\hat{\bf d})\cdot(\partial_{k_b}\hat{\bf d})\right], \label{eq:QM_ab}
\end{eqnarray} where $\hat{\bf d}\!=\!(d_{0}/d,d_{x}/d,d_{y}/d,d_{z}/d)$ with $d\!=\!(d_{x}^{2}\!+\!d_{y}^{2}\!+\!d_{z}^{2})^{1/2}\!=\!\sqrt{(i\gamma\!+\!M\!-\!\alpha k^{2})^{2}\!+\!v^{2}k^{2}}$. Derivation details of Eq.~\eqref{eq:QM_ab} can be found in Appendix~\ref{Appendix_7}.

\section{Nonlinear responses}\label{section4}

In this section, we will present a detailed analysis of both the extrinsic and intrinsic nonlinear responses.

\subsection{Berry curvature dipole and extrinsic nonlinear response}\label{section4.2}

The BCD contributes to the higher-order effects in the extrinsic nonlinear response. To calculate the BCD [Eq.~\eqref{eq:BCD}] in our exceptional ring system, a key term is the momentum gradient of the non-Hermitian Berry curvature [Eq.~\eqref{eq:BC}] with respect to $k_{x}$, which is given by 
\begin{eqnarray}
P_{yxx}^{(\pm)}&\!\equiv\!&\frac{\partial\Omega_{z}^{(\pm)}}{\partial k_x} \nonumber\\
&\!=\!& 
\pm{\rm Re}\!\left\{\!\!\frac{\eta k_{x}v^{2} \!\left[2\alpha\left(g_{-}^{2}\!+\!v^{2}k^{2}\right) \!-\! 3g_{+}^{}\left(v^{2} \!-\! 2\alpha g_{-}^{}\right)\right]}{2\left(g_{-}^{2}\!+\!v^{2}k^{2}\right)^{5/2}}\!\!\right\}\!,\label{eq:BC_kx}\nonumber\\
\end{eqnarray} where $g_{\pm}^{}\!=\!i\gamma \!+\! M \!\pm\! \alpha k^{2}$ and $\partial_{k_x}\Omega_{z}^{(\pm)}(-k_{x})\!=\!-\partial_{k_x}\Omega_{z}^{(\pm)}(k_{x})$.

In our work, we consider zero temperature, such that the corresponding equilibrium Fermi-Dirac distribution function can be expressed as $f_{0}^{(\pm)}\!=\!f_{0}^{(\pm)}\left[{\rm Re}\left(\epsilon_{\bf k}^{(\pm)}\right)\!-\!E_{F}\right]\!=\!\Theta\left[E_{F}\!-\!{\rm Re}\left(\epsilon_{\bf k}^{(\pm)}\right)\right]$, with the Fermi energy $E_{F}$ and the Heaviside function $\Theta(x)$~\cite{qin2015three,qin2018high,qin2019polaron,qin2018universal,qin2017width}. For simplicity, we choose the parameters that satisfy the condition $\sum_{\pm}[f_{0}^{(\pm)}\Omega_{z}^{(\pm)}]\big|_{k_x=-\pi}^{k_x=\pi}\!=\!0$. As such, the first term in the BCD [Eq.~\eqref{eq:BCD}] vanishes. To achieve this, we need a significant energy gap at $k_{x}\!=\!\pm\pi$, as illustrated in Fig.~\ref{fig:E3D_t02_g1_M5}(a), i.e., when $k_{x}\!=\!\pm\pi$ is fixed, a significant gap naturally appears in the 2D slice energy spectrum as a function of $k_{y}$, as shown in Fig.~\ref{fig:E3D_t02_g1_M5}(a). Given that the Berry curvature [Eq.~\eqref{eq:BC}] is even about $k_{x}$, we then have $\Omega_{z}^{(\pm)}(k_{x}\!=\!\pi)\!=\!\Omega_{z}^{(\pm)}(k_{x}\!=\!-\pi)$. 
To achieve nonvanishing BCD, we need the conditions: $\Theta\left[E_{F}\!-\!{\rm Re}\left(\epsilon_{k_{x}=\pi}^{(\pm)}\right)\right]\!=\!\Theta\left[E_{F}\!-\!{\rm Re}\left(\epsilon_{k_{x}=-\pi}^{(\pm)}\right)\right]$ 
and  ${\rm Re}\left(\epsilon_{k_{x}=\pm\pi}^{(-)}\right)<E_{F}<{\rm Re}\left(\epsilon_{k_{x}=\pm\pi}^{(+)}\right)~(k_{y}\in[-\pi,\pi])$ as shown in Fig.~\ref{fig:E3D_t02_g1_M5}(a). For our model, the BCD [Eq.~\eqref{eq:BCD}] hence does not vanish only at nonzero tilt ($t_{0}\!\neq\!0$).

\begin{figure}[htpb]
\centering
\includegraphics[width=0.7\columnwidth]{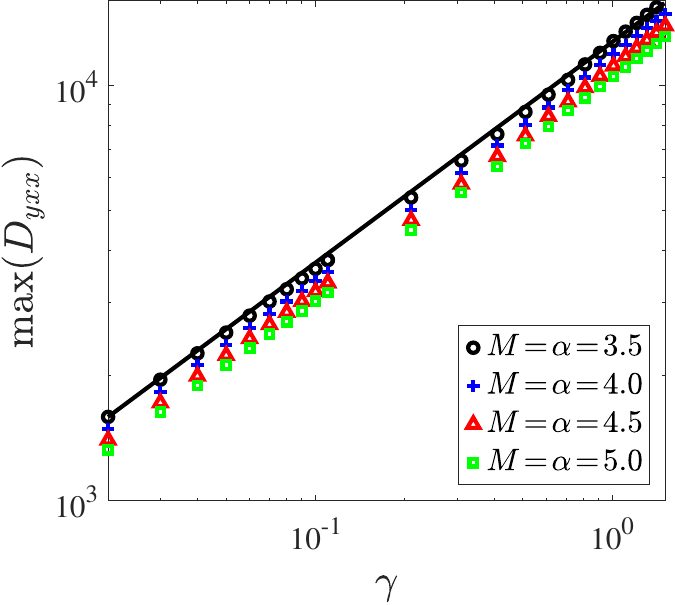}
\caption{\textbf{Maximum value of the extrinsic nonlinear response of an exceptional ring as a function of $\gamma$, for fixed radius $R\!=\!1$.} We set $\gamma\!=\!v$ and  $M=\alpha$ such that $R\!=\!1$, and numerically plot the maximum value of the Berry curvature dipole (BCD) [Eq.~\eqref{eq:BCD}]. For different values of $M\!=\!\alpha\!=\!3.5$ (black circle), 4.0 (blue cross), 4.5 (red triangle), and 5.0 (green square), the numerical data follows the same trend as the solid line, which is fitted from the ansatz ${\rm Max}(D_{yxx})\!=\!\exp(a_{0}\ln\gamma\!+\!b_{0})$ ($a_{0}\!=\!0.5313$ and $b_{0}\!=\!9.4517$), demonstrating an approximate but universal power scaling law ${\rm Max}(D_{yxx})\!\propto\!\sqrt{\gamma}$. The other parameters are the same as those in Fig.~\ref{fig:BCD_Pxz_t02_g_v_together_M5}. 
}\label{fig:BCD_max_t02_g_v_M_analytic_together}
\end{figure}

According to Eq.~\eqref{eq:BCD}, the predominant contribution to the nonlinear response stems from the states near the Fermi surface. Specifically, when the Fermi surface approaches the exceptional ring, these exceptional states can give rise to an intense response.  In Figs.~\ref{fig:BCD_Pxz_t02_g_v_together_M5}(a1)-\ref{fig:BCD_Pxz_t02_g_v_together_M5}(d1), we plot the BCD near the exceptional ring ($R\!=\!1$) with the Fermi level $E_{F}\in[-0.3,0.3]$.  Here, strong extrinsic nonlinear BCD responses are evident, highlighted by the two prominent peaks, particularly when the loss parameter $\gamma$  increases from $0.1$ to $1.5$.  
Actually, the peaks appear only in Fig.~\ref{fig:BCD_Pxz_t02_g_v_together_M5}(a1) for the very small parameter value of $\gamma\!=\!0.1$. However, as $\gamma$ increases, the extrinsic nonlinear Hall response develops approximate plateaus, as seen in Figs.~\ref{fig:BCD_Pxz_t02_g_v_together_M5}(b1)-\ref{fig:BCD_Pxz_t02_g_v_together_M5}(d1). Moreover, the width of these plateaus grows with $\gamma$, indicating a broadening effect as the parameter increases.

Next, we examine the BCD density profile $P_{yxx}^{(\pm)}({\bf k})$ itself.
As depicted in Figs.~\ref{fig:BCD_Pxz_t02_g_v_together_M5}(a2)-\ref{fig:BCD_Pxz_t02_g_v_together_M5}(d2), near the exceptional ring where bands coalesce at $k^{2}\!=\!R^{2}\!=\!M/\alpha$, the BCD density profile markedly diverges. The details for the scaling behavior of the BCD density profile near the exceptional ring can be found in Appendix~\ref{Appendix_8}.

To understand how the BCD response scales with the non-Hermiticity for an exceptional ring at fixed radius (we set it to $R\!=\!1$), we evaluate the maximum value of the BCD as a function of $\gamma$ for different parameters $M\!=\!\alpha$, while keeping the radius $R\!=\!1$ fixed. As shown in Fig.~\ref{fig:BCD_max_t02_g_v_M_analytic_together}, the maximum BCD value increases like $\sqrt{\gamma}$ as non-Hermiticity $\gamma$ increases, indicating that non-Hermiticity enhances the extrinsic nonlinear Hall effect.
To understand this, we then consider the condition of $0\ll|\gamma|\!=\!|v|\ll\alpha\!=\!M$ and expand Eq.~\eqref{eq:BC_kx} near the exceptional ring along $k\!=\!\sqrt{M/\alpha}\!+\!\delta k\!=\!1\!+\!\delta k$ and $k_{x}\!=\!(1\!+\!\delta k)\cos\theta$ with $M\!=\!\alpha$ and $\delta k\to0$ as follows:
\begin{eqnarray}
\frac{\partial\Omega_{z}^{(\pm)}}{\partial k_x} 
\!\propto\!(v/\alpha)^{1/2}\!\propto\!\sqrt{\gamma}.\label{eq:BCD_v_S4} 
\end{eqnarray} Details on the derivation of Eq.~\eqref{eq:BCD_v_S4} can be found in Appendix~\ref{Appendix_9}.
Therefore, we obtain the derived analytical scaling behavior $\sqrt{\gamma}$ described by Eq.~\eqref{eq:BCD_v_S4}, which aligns with our numerical results shown in Fig.~\ref{fig:BCD_max_t02_g_v_M_analytic_together}. However, this approximation is not entirely accurate if $v$ is exceedingly small in the initial $\delta k$ approximation. Numerically, the smallest $\delta k$ is determined by the resolution of the $k$ intervals, which in turn depends on the physical size of the system. For finite lattices, the maximum $D_{yxx}$ scales as $\sqrt{\gamma}$ down to the spatial scale of lattices.

From Fig.~\ref{fig:BCD_max_t02_g_v_M_analytic_together}, the maximum of the BCD decreases significantly as $\gamma\!=\!v\!\to\!0$. In fact, it tends to zero, as can be seen by expanding the term $\partial\Omega_{z}^{(\pm)}/\partial k_{x}$ [Eq.~\eqref{eq:BC_kx}] in the BCD as follows:
\begin{eqnarray}
\lim_{\gamma=v\to0}\frac{\partial\Omega_{z}^{(\pm)}}{\partial k_x} 
\!\approx\! \!\pm\lim_{\gamma\to0}{\rm Re}\!\left[\!\frac{2\alpha k_{x}\gamma^{2}(2\!+\!k^{2})}{M^{3}(1\!-\!k^{2})^{4}}\!\right]\!.\label{eq:BC_kx_v0}
\end{eqnarray} Details on the derivation of Eq.~\eqref{eq:BC_kx_v0} can be found in Appendix~\ref{Appendix_10}.
The approximation Eq.~\eqref{eq:BC_kx_v0} shows that $\lim_{\gamma=v\to0}D_{yxx}\!=\!0$, which is consistent with our numerical results in Fig.~\ref{fig:BCD_max_t02_g_v_M_analytic_together}.

\subsection{Berry connection polarizability and intrinsic nonlinear response}\label{section4.3}

Here, we examine the intrinsic nonlinear response in our exceptional ring model~\eqref{eq:H}. The $yxx$-component intrinsic nonlinear velocity response $\bar{v}_{yxx}^{\rm BCP}$ [Eq.~\eqref{eq:BCP_abc}] under the BCP is given by~\cite{gao2014field,liu2021intrinsic,wang2021intrinsic,liu2022berry}
\begin{eqnarray}
\bar{v}_{yxx}^{\rm BCP}
&\!=\!&\sum_{n,n\rq{}=+,-}^{\epsilon_{\bf k}^{(n)}\neq \epsilon_{\bf k}^{(n\rq{})}}\!\Gamma_{xx}\!\int\frac{dk_x}{2\pi}\!\left[\frac{f_{0}^{(n)}{\cal G}_{xx}^{(n)}}{\epsilon_{\bf k}^{(n)}-\epsilon_{\bf k}^{(n\rq{})}}\right]\bigg|_{k_y=-\pi}^{k_y=\pi}  \nonumber\\
&&\!+\! \sum_{n,n\rq{}=+,-}^{\epsilon_{\bf k}^{(n)}\neq \epsilon_{\bf k}^{(n\rq{})}}\!\Gamma_{xx}\!\int\frac{dk_y}{2\pi}\!\left[\frac{-f_{0}^{(n)}{\cal G}_{yx}^{(n)}}{\epsilon_{\bf k}^{(n)}-\epsilon_{\bf k}^{(n\rq{})}}\right]\bigg|_{k_x=-\pi}^{k_x=\pi} \nonumber\\
&&\!-\!\!\sum_{n,n\rq{}=+,-}^{\epsilon_{\bf k}^{(n)}\neq \epsilon_{\bf k}^{(n\rq{})}}\!\Gamma_{xx}\!\int\!\!\frac{d^{2}\mathbf{k}}{(2\pi)^{2}}f_{0}^{(n)}\!\!\left[\!\frac{\partial_{k_y}{\cal G}_{xx}^{(n)}\!-\!\partial_{k_x}{\cal G}_{yx}^{(n)}}{\epsilon_{\bf k}^{(n)}-\epsilon_{\bf k}^{(n\rq{})}}\!\right]\!\! \nonumber\\
&&\!+\! \sum_{n,n\rq{}=+,-}^{\epsilon_{\bf k}^{(n)}\neq \epsilon_{\bf k}^{(n\rq{})}}\!\Gamma_{xx}\!\int\!\!\frac{d^{2}\mathbf{k}}{(2\pi)^{2}}\!\!\frac{f_{0}^{(n)}}{\left(\epsilon_{\bf k}^{(n)}\!-\!\epsilon_{\bf k}^{(n\rq{})}\right)^{2}} \nonumber\\
&&\times\left[{\cal G}_{xx}^{(n)}\!\partial_{k_y}\!\!\left(\epsilon_{\bf k}^{(n)}\!-\!\epsilon_{\bf k}^{(n\rq{})}\right) \!-\! {\cal G}_{yx}^{(n)}\!\partial_{k_x}\!\!\left(\epsilon_{\bf k}^{(n)}\!-\!\epsilon_{\bf k}^{(n\rq{})}\right)\right], \label{eq:BCP_yxx} \nonumber\\
\end{eqnarray} 
where $\Gamma_{xx}\!=\!2{\cal S}\mathcal{F}_{x}^{2}/\hbar$ with $\mathcal{F}_{x}$ being the magnitude of the applied force, $v_{a}^{(n)}\!=\!\partial\epsilon_{\bf k}^{(n)}/\partial k_{a}$ the group velocity of the $n$th band~\cite{zhuang2024intrinsic}, and we can have
\begin{eqnarray}
\frac{v_{y}^{(\pm)}\!-\!v_{y}^{(\mp)}}{\left(\epsilon_{\bf k}^{(\pm)}\!-\!\epsilon_{\bf k}^{(\mp)}\right)^{2}}&\!=\!&\pm\frac{k_{y}[v^{2}\!-\!2\alpha(i\gamma\!+\!M\!-\!\alpha k^{2})]}{2d^{3}}, \label{eq:vyy}\\
\frac{v_{x}^{(\pm)}\!-\!v_{x}^{(\mp)}}{\left(\epsilon_{\bf k}^{(\pm)}\!-\!\epsilon_{\bf k}^{(\mp)}\right)^{2}}&\!=\!&\pm\frac{k_{x}[v^{2}\!-\!2\alpha(i\gamma\!+\!M\!-\!\alpha k^{2})]}{2d^{3}}, \label{eq:vxx}
\end{eqnarray} Here we have written $d\!=\!(d_{x}^{2}\!+\!d_{y}^{2}\!+\!d_{z}^{2})^{1/2}\!=\!\sqrt{(i\gamma\!+\!M\!-\!\alpha k^{2})^{2}\!+\!v^{2}k^{2}}$. The quantum metric components are given by ($\eta\!=\!-1$)
\begin{eqnarray}
{\cal G}_{xx}^{(\pm)}
&\!=\!&{\rm Re}\!\left\{\!\frac{v^{2}}{4d^{2}}\left\{1\!+\!\frac{k_{x}^{2}[4\alpha(M\!+\!i\gamma) \!-\! v^{2}]}{d^{2}} \right\}\!\right\}\!, \label{eq:Gxx} \\
{\cal G}_{yx}^{(\pm)}
&\!=\!&{\rm Re}\!\left\{\!\!\frac{k_{x}k_{y}v^{2}[4\alpha(M\!+\!i\gamma) \!-\! v^{2}]}{4d^{4}}\!\!\right\}\!.\label{eq:Gyx}
\end{eqnarray}
Furthermore, we can obtain
\begin{eqnarray}
\frac{\partial{\cal G}_{xx}^{(\pm)}}{\partial k_y}
&\!=\!& \!{\rm Re}\!\Bigg\{\!\frac{k_{y}v^{2}(2\alpha g_{-}^{}\!-\!v^{2})}{2\left(g_{-}^{2}\!+\!v^{2}k^{2}\right)^{3}} \nonumber\\
&&\!\times\!\left\{\!\left(g_{-}^{2}\!+\!v^{2}k^{2}\right)\!+\!2k_{x}^{2}[4\alpha(M\!+\!i\gamma) \!-\! v^{2}]\right\}\!\!\!\Bigg\}\!,\label{eq:Gxxky} \\
\frac{\partial{\cal G}_{yx}^{(\pm)}}{\partial k_x}
&\!=\!& \!{\rm Re}\!\Bigg\{\!\frac{k_{y}v^{2}[4\alpha(M\!+\!i\gamma)\!-\!v^{2}]}{4\left(g_{-}^{2}\!+\!v^{2}k^{2}\right)^{3}} \nonumber\\
&&\!\times\!\left[\left(g_{-}^{2}\!+\!v^{2}k^{2}\right)\!+\!4k_{x}^{2}(2\alpha g_{-}^{}\!-\!v^{2})\right]\!\Bigg\}\!,\label{eq:Gyxkx}
\end{eqnarray} 
where $g_{-}^{}\!=\!i\gamma \!+\! M \!-\! \alpha k^{2}$. 

\begin{figure}
\centering
\includegraphics[width=0.7\columnwidth]{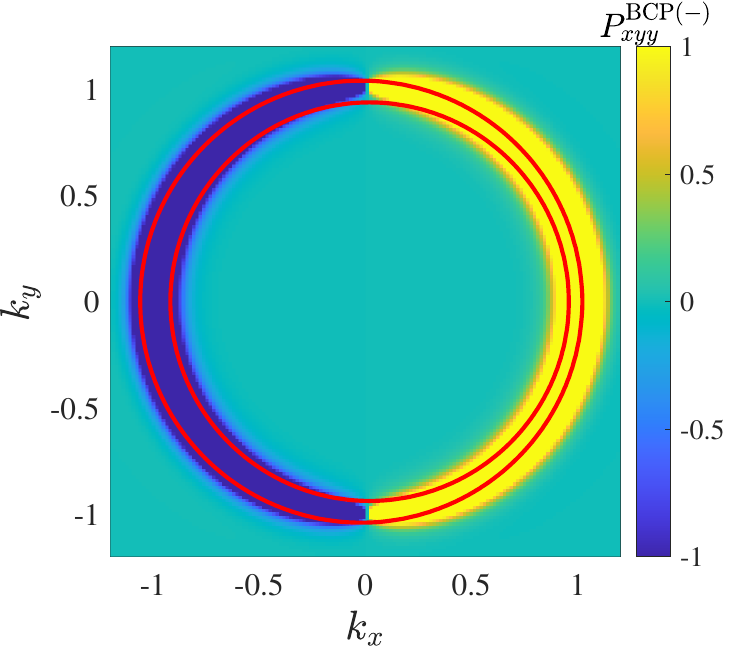}
\caption{\textbf{Fermi surface (red curves)} of our exceptional ring model [Eq.~\eqref{eq:H}] with model parameters $t_{0}\!=\!0.2$, $\gamma\!=\!v\!=\!1$, and $M\!=\!\alpha\!=\!5$, correspond to a ring radius $R\!=\!1$. The Fermi region is slightly broadened due to the nonzero Fermi energy of $E_{F}\!=\!0.5$, and is asymmetric about $k_{x}$ but symmetric about $k_{y}$. The color denotes the BCP density profile $P_{xyy}^{{\rm BCP}(-)}({\bf k})$ defined in Eq.~\eqref{eq:BCP_v}.
}\label{fig:FermiSurface}
\end{figure}

First, the integrand $[f_{0}^{(\pm)}{\cal G}_{xx}^{(\pm)}/(\epsilon_{\bf k}^{(\pm)}\!-\!\epsilon_{\bf k}^{(\mp)})]\big|_{k_y=-\pi}^{k_y=\pi}$ in Eq.~\eqref{eq:BCP_yxx} is even about $k_{y}$ and $\epsilon_{k_{y}=\pi}^{(\pm)}\!=\!\epsilon_{k_{y}=-\pi}^{(\pm)}$, so that it vanishes after the integration over the Fermi surface (as shown in Fig.~\ref{fig:FermiSurface}). Figure~\ref{fig:FermiSurface} shows that the Fermi surface of our tilted massive 2D Dirac model [Eq.~\eqref{eq:H}] is asymmetric about $k_{x}$ but symmetric about $k_{y}$. Second, the integrand $[-f_{0}^{(\pm)}{\cal G}_{yx}^{(\pm)}/(\epsilon_{\bf k}^{(\pm)}\!-\!\epsilon_{\bf k}^{(\mp)})]\big|_{k_x=-\pi}^{k_x=\pi}$ in Eq.~\eqref{eq:BCP_yxx} is odd about $k_{y}$ so that it also vanishes after the integration over the Fermi surface.
Indeed, these symmetry consideration dictate the vanish of some BCP components. From Eqs.~\eqref{eq:Gxxky} and \eqref{eq:Gyxkx}, the integrand $[\partial_{k_y}{\cal G}_{xx}^{(\pm)}\!-\!\partial_{k_x}{\cal G}_{yx}^{(\pm)}]/[\epsilon_{\bf k}^{(\pm)}\!-\!\epsilon_{\bf k}^{(\mp)}]$ is odd about $k_{y}$ and even about $k_{x}$, leading to the vanishing of the third term of Eq.~\eqref{eq:BCP_yxx} after the integration over the Fermi surface. Similarly, from Eqs.~\eqref{eq:vyy}, \eqref{eq:vxx}, \eqref{eq:Gxx}, and \eqref{eq:Gyx}, the integrand $[{\cal G}_{xx}^{(\pm)}(v_{y}^{(\pm)}\!-\!v_{y}^{(\mp)})\!-\!{\cal G}_{yx}^{(\pm)}(v_{x}^{(\pm)}\!-\!v_{x}^{(\mp)})]/[\epsilon_{\bf k}^{(\pm)}\!-\!\epsilon_{\bf k}^{(\mp)}]^{2}$ is odd about $k_{y}$ and even about $k_{x}$, leading to the vanishing of the fourth term of Eq.~\eqref{eq:BCP_yxx} after the integration over the Fermi surface. Therefore, we obtain $\bar{v}_{yxx}^{\rm BCP}=0$. In the same way, we can find that the $xyx$-component $\bar{v}_{xyx}^{\rm BCP}\!=\!-\bar{v}_{yxx}^{\rm BCP}$ equals zero too. Besides, we also have $\bar{v}_{xxy}^{\rm BCP}\!=\!0$.

\begin{figure*}[htpb]
\centering
\includegraphics[width=.8\textwidth]{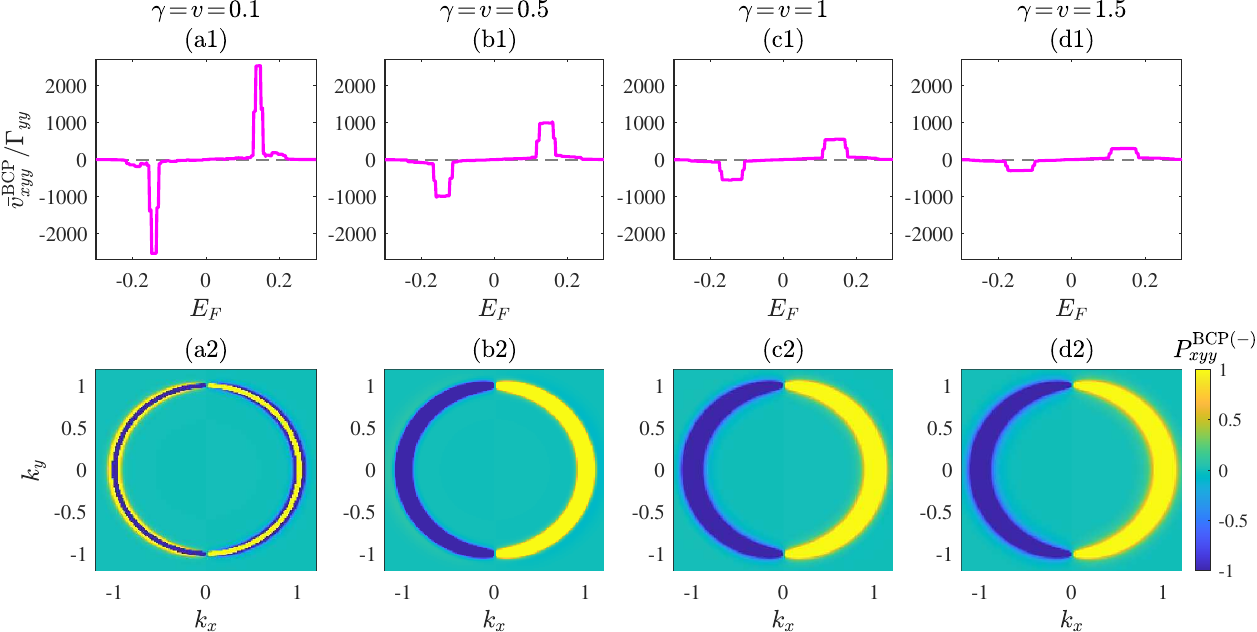}
\caption{\textbf{Intrinsic nonlinear response with different values of $\gamma\!=\!v$ for fixed radius $R\!=\!\gamma/v\!=\!\sqrt{M/\alpha}\!=\!1$ with $M\!=\!\alpha\!=\!5$.} (a1-d1): The intrinsic nonlinear velocity $\bar{v}_{xyy}^{\rm BCP}$ [Eq.~\eqref{eq:BCP_xyy}] for the Berry connection polarizability, which is suppressed with increasing non-Hermiticity. (a2)-(d2): The BCP density profile $P_{xyy}^{{\rm BCP}(-)}({\bf k})$ [Eq.~\eqref{eq:BCP_v}] in the $k_{x}$--$k_{y}$ plane. The other parameter is $t_{0}\!=\!0.2$.
}
\label{fig:BCP_P_t02_g_v_together_M5}
\end{figure*}

\begin{figure}[htpb]
\centering
\includegraphics[width=0.7\columnwidth]{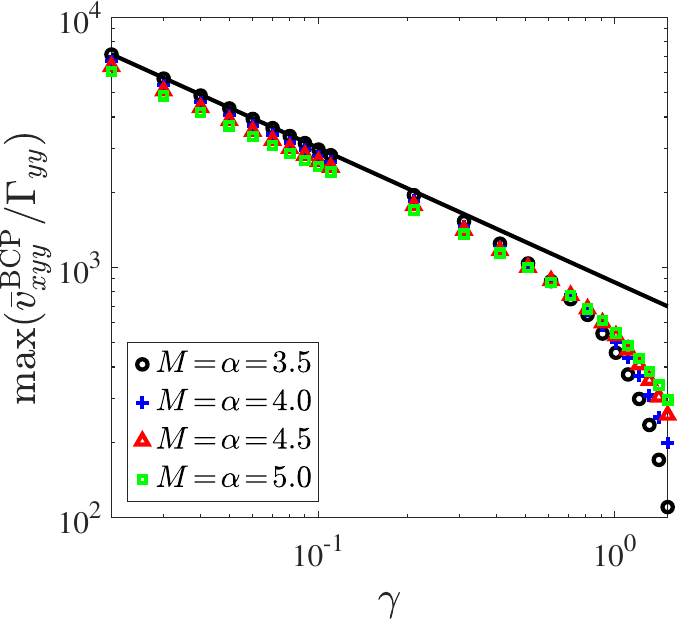}
\caption{\textbf{Peak of the intrinsic nonlinear response $\bar{v}_{xyy}^{\rm BCP}$ as a function of non-Hermiticity $\gamma\!=\!v$.} We plot the maximum value of $\bar{v}_{xyy}^{\rm BCP}$ [Eq.~\eqref{eq:BCP_xyy}] with $\gamma\!=\!v$, such that $R\!=\!1$. Results are similar across different values of $M\!=\!\alpha\!=\!3.5$ (black circle), 4.0 (blue cross), 4.5 (red trangle), and 5.0 (green square). The The solid line show the power-law dependence of the maximum $\bar{v}_{xyy}^{\rm BCP}$, which is fitted from the ansatz ${\rm Max}(\bar{v}_{xyy}^{\rm BCP}/\Gamma_{yy})\!=\!\exp(a_{1}\ln\gamma+b_{1})$ in the small $\gamma$ regime. For the case $M\!=\!\alpha\!=\!3.5$, we have $a_{1}\!=\!-0.5370$, $b_{1}\!=\!6.7711$ for the regime $\gamma\!=\!v\in[0.02,0.21]$, approximately obeying the scaling law ${\rm Max}(\bar{v}_{xyy}^{\rm BCP}/\Gamma_{yy})\!\propto\!\sqrt{1/\gamma}$. However, in the large $\gamma$ regime, there is no such power-law scaling. The other parameters are the same as those in Fig.~\ref{fig:BCP_P_t02_g_v_together_M5}. 
}\label{fig:BCP_max_t02_g_v_M_analytic_together}
\end{figure}

One nonvanishing component in the intrinsic nonlinear response is the $xyy$-component intrinsic nonlinear velocity $\bar{v}_{xyy}^{\rm BCP}$ [Eq.~\eqref{eq:BCP_abc}]: 
\begin{eqnarray}
\bar{v}_{xyy}^{\rm BCP}
&\!=\!&\sum_{n,n\rq{}=+,-}^{\epsilon_{\bf k}^{(n)}\neq \epsilon_{\bf k}^{(n\rq{})}}\!\Gamma_{yy}\!\int\frac{dk_y}{2\pi}\!\left[\frac{f_{0}^{(n)}{\cal G}_{yy}^{(n)}}{\epsilon_{\bf k}^{(n)}-\epsilon_{\bf k}^{(n\rq{})}}\right]\bigg|_{k_x=-\pi}^{k_x=\pi}  \nonumber\\
&&\!+\! \sum_{n,n\rq{}=+,-}^{\epsilon_{\bf k}^{(n)}\neq \epsilon_{\bf k}^{(n\rq{})}}\!\Gamma_{yy}\!\int\frac{dk_x}{2\pi}\!\left[\frac{-f_{0}^{(n)}{\cal G}_{xy}^{(n)}}{\epsilon_{\bf k}^{(n)}-\epsilon_{\bf k}^{(n\rq{})}}\right]\bigg|_{k_y=-\pi}^{k_y=\pi} \nonumber\\
&&\!-\!\!\sum_{n,n\rq{}=+,-}^{\epsilon_{\bf k}^{(n)}\neq \epsilon_{\bf k}^{(n\rq{})}}\!\Gamma_{yy}\!\int\!\!\frac{d^{2}\mathbf{k}}{(2\pi)^{2}}f_{0}^{(n)}\!\!\left[\!\frac{\partial_{k_x}{\cal G}_{yy}^{(n)}\!-\!\partial_{k_y}{\cal G}_{xy}^{(n)}}{\epsilon_{\bf k}^{(n)}-\epsilon_{\bf k}^{(n\rq{})}}\!\right]\!\! \nonumber\\
&&\!+\! \sum_{n,n\rq{}=+,-}^{\epsilon_{\bf k}^{(n)}\neq \epsilon_{\bf k}^{(n\rq{})}}\!\Gamma_{yy}\!\int\!\!\frac{d^{2}\mathbf{k}}{(2\pi)^{2}}\!\!\frac{f_{0}^{(n)}}{\left(\epsilon_{\bf k}^{(n)}\!-\!\epsilon_{\bf k}^{(n\rq{})}\right)^{2}} \nonumber\\
&&\times\left[{\cal G}_{yy}^{(n)}\!\partial_{k_x}\!\!\left(\epsilon_{\bf k}^{(n)}\!-\!\epsilon_{\bf k}^{(n\rq{})}\right) \!-\! {\cal G}_{xy}^{(n)}\!\partial_{k_y}\!\!\left(\epsilon_{\bf k}^{(n)}\!-\!\epsilon_{\bf k}^{(n\rq{})}\right)\right], \label{eq:BCP_xyy} \nonumber\\
\end{eqnarray} where $\Gamma_{yy}\!=\!2{\cal S}\mathcal{F}_{y}^{2}/\hbar$ with the magnitude of force $\mathcal{F}_{y}$, and the quantum metric components are ($\eta\!=\!-1$)
\begin{eqnarray}
{\cal G}_{yy}^{(\pm)}
&\!=\!&{\rm Re}\!\left\{\!\frac{v^{2}}{4d^{2}}\left\{1\!+\!\frac{k_{y}^{2}[4\alpha(M\!+\!i\gamma) \!-\! v^{2}]}{d^{2}} \right\}\!\!\right\}\!, \label{eq:Gyy}\\
{\cal G}_{xy}^{(\pm)}
&\!=\!&{\rm Re}\!\left\{\!\!\frac{k_{x}k_{y}v^{2}[4\alpha(M\!+\!i\gamma) \!-\! v^{2}]}{4d^{4}}\!\!\right\}\!,\label{eq:Gxy}
\end{eqnarray} 
giving rise to momentum-space derivatives  
\begin{eqnarray}
\frac{\partial{\cal G}_{yy}^{(\pm)}}{\partial k_x}
&\!=\!& \!{\rm Re}\!\Bigg\{\!\frac{k_{x}v^{2}(2\alpha g_{-}^{} \!-\! v^{2})}{2\left(g_{-}^{2}\!+\!v^{2}k^{2}\right)^{3}} \nonumber\\
&&\!\times\!\left\{\!\left(g_{-}^{2}\!+\!v^{2}k^{2}\right)\!+\!2k_{y}^{2}[4\alpha(M \!+\! i\gamma) \!-\! v^{2}] \right\}\!\!\!\Bigg\}\!,\label{eq:Gyykx}\\
\frac{\partial{\cal G}_{xy}^{(\pm)}}{\partial k_y}
&\!=\!& \!{\rm Re}\!\Bigg\{\!\frac{k_{x}v^{2}[4\alpha(M\!+\!i\gamma)\!-\!v^{2}]}{4\left(g_{-}^{2}\!+\!v^{2}k^{2}\right)^{3}} \nonumber\\
&&\!\times\![\left(g_{-}^{2}\!+\!v^{2}k^{2}\right)\!+\!4k_{y}^{2}(2\alpha g_{-}^{}\!-\!v^{2})]\!\Bigg\}\!.\label{eq:Gxyky}
\end{eqnarray}

Now, the integrand $\left[f_{0}^{(\pm)}{\cal G}_{yy}^{(\pm)}/(\epsilon_{\bf k}^{(\pm)}\!-\!\epsilon_{\bf k}^{(\mp)})\right]\Big|_{k_y=-\pi}^{k_y=\pi}$ in Eq.~\eqref{eq:BCP_xyy} is even about $k_{x}$ and $\Theta\left[E_{F}\!-\!{\rm Re}\left(\epsilon_{k_{x}=\pi}^{(\pm)}\right)\right]\!=\!\Theta\left[E_{F}\!-\!{\rm Re}\left(\epsilon_{k_{x}=-\pi}^{(\pm)}\right)\right]$ under present parameters as shown in Fig.~\ref{fig:E3D_t02_g1_M5}(a). Under the current parameters illustrated in Fig.~\ref{fig:E3D_t02_g1_M5}(a), a significant energy gap is present to satisfy this condition, so that it vanishes after the integration over the Fermi surface. Second, the integrand $\sum_{n,n\rq{}=+,-}\left[-f_{0}^{(n)}{\cal G}_{xy}^{(n)}/(\epsilon_{\bf k}^{(n)}\!-\!\epsilon_{\bf k}^{(n\rq{})})\right]\Big|_{k_y=-\pi}^{k_y=\pi}$ in Eq.~\eqref{eq:BCP_xyy} vanishes after the summation of the band index $n,n\rq{}=+,-$ with the condition $\epsilon_{k_{y}=\pi}^{(\pm)}\!=\!\epsilon_{k_{y}=-\pi}^{(\pm)}$.
From Eqs.~\eqref{eq:Gyykx} and \eqref{eq:Gxyky}, the integrand $[\partial_{k_x}{\cal G}_{yy}^{(\pm)}\!-\!\partial_{k_y}{\cal G}_{xy}^{(\pm)}]/[\epsilon_{\bf k}^{(\pm)}\!-\!\epsilon_{\bf k}^{(\mp)}]$ is odd about $k_{x}$ and even about $k_{y}$, hence a nonzero $\bar{v}_{xyy}^{\rm BCP}$ is permitted after the integration over the Fermi surface. Similarly, from Eqs.~\eqref{eq:vyy}, \eqref{eq:vxx}, \eqref{eq:Gyy}, and \eqref{eq:Gxy}, the integrand $[{\cal G}_{yy}^{(\pm)}(v_{x}^{(\pm)}\!-\!v_{x}^{(\mp)})\!-\!{\cal G}_{xy}^{(\pm)}(v_{y}^{(\pm)}\!-\!v_{y}^{(\mp)})]/[\epsilon_{\bf k}^{(\pm)}\!-\!\epsilon_{\bf k}^{(\mp)}]^{2}$ is odd about $k_{x}$ and even about $k_{y}$, leading to the nonvanishing of the fourth term of Eq.~\eqref{eq:BCP_xyy} after the integration over the Fermi surface. Moreover, we have $\bar{v}_{yxy}^{\rm BCP}\!=\!-\bar{v}_{xyy}^{\rm BCP}\!\neq\!0$ and $\bar{v}_{yyx}^{\rm BCP}\!=\!0$. Therefore, we only need to focus on the nonzero component $\bar{v}_{xyy}^{\rm BCP}$.

In Figs.~\ref{fig:BCP_P_t02_g_v_together_M5}(a1)-\ref{fig:BCP_P_t02_g_v_together_M5}(d1), we plot the intrinsic nonlinear velocity $\bar{v}_{xyy}^{\rm BCP}$ [Eq.~\eqref{eq:BCP_xyy}] under different values of the tunable effective gain/loss $\gamma$ at fixed radius $R\!=\!1$ of the exceptional ring. Similar to the BCD, $\bar{v}_{xyy}^{\rm BCP}$ exhibits a pair of peaks within the Fermi energy range $E_{F}\in[-0.3,0.3]$. 
Actually, the peaks occur only in Fig.~\ref{fig:BCD_Pxz_t02_g_v_together_M5}(a1) and Fig.~\ref{fig:BCP_P_t02_g_v_together_M5}(a1) for the very small parameter $\gamma\!=\!0.1$. However, for larger values of $\gamma$, we observe approximate plateaus in the nonlinear Hall responses, as shown in Figs.~\ref{fig:BCD_Pxz_t02_g_v_together_M5}(b1)-\ref{fig:BCD_Pxz_t02_g_v_together_M5}(d1) and Figs.~\ref{fig:BCP_P_t02_g_v_together_M5}(b1)-\ref{fig:BCP_P_t02_g_v_together_M5}(d1). Furthermore, the width of these approximate plateaus increases with $\gamma$, indicating a $\gamma$-dependent behavior. To elucidate the origin of the approximate plateaus in the nonlinear Hall responses as a function of the Fermi energy $E_{F}$ as shown in Figs.~\ref{fig:BCD_Pxz_t02_g_v_together_M5}(a1)-\ref{fig:BCD_Pxz_t02_g_v_together_M5}(d1) and Figs.~\ref{fig:BCP_P_t02_g_v_together_M5}(a1)-\ref{fig:BCP_P_t02_g_v_together_M5}(d1), we analyze the $k_{y}\!=\!0$ slice of the real part of the energy band structure for $\gamma\!=\!1$ as shown in Fig.~\ref{fig:E3D_t02_g1_M5}(b). Figure~\ref{fig:E3D_t02_g1_M5}(b) reveals that the upper and lower bands merge and become linear within a narrow gap region near the exceptional point. Two energy gaps are symmetrically positioned about the Re($\epsilon_{\bf k}^{})\!=\!0$ axis. Importantly, the approximate plateaus in the nonlinear Hall responses correspond to these small energy gaps.
However, in contrast to the BCD, Fig.~\ref{fig:BCP_max_t02_g_v_M_analytic_together} illustrates that the intrinsic nonlinear Hall effect is suppressed as non-Hermiticity increases. Substituting Eqs.~\eqref{eq:Gyykx} and \eqref{eq:Gxyky} into the integrand of $\bar{v}_{xyy}^{\rm BCP}$, one can obtain
\begin{eqnarray}
P_{xyy}^{{\rm BCP}(\pm)}({\bf k})&\!\equiv\!&\!-\!\!\left[\!\frac{\partial_{k_x}{\cal G}_{yy}^{(\pm)}\!-\!\partial_{k_y}{\cal G}_{xy}^{(\pm)}}{\epsilon_{\bf k}^{(\pm)}-\epsilon_{\bf k}^{(\mp)}} \right]\! \nonumber\\
&&\!+\!\left[\frac{{\cal G}_{yy}^{(\pm)}(v_{x}^{(\pm)}\!-\!v_{x}^{(\mp)})\!-\!{\cal G}_{xy}^{(\pm)}(v_{y}^{(\pm)}\!-\!v_{y}^{(\mp)})}{\left(\epsilon_{\bf k}^{(\pm)}\!-\!\epsilon_{\bf k}^{(\mp)}\right)^{2}} \right]\! \nonumber\\
&\!=\!&\pm{\rm Re}\!\left\{\!\frac{k_{x}v^{2}\left[v^{2} \!-\! \alpha\left(M \!-\! 3\alpha^{2}k^{2} \!+\! i\gamma\right) \right]}{4(g_{-}^{2} \!+\! v^{2}k^{2})^{5/2}} \!\right\}\!, \label{eq:BCP_v}\nonumber\\
\end{eqnarray} where $P_{xyy}^{{\rm BCP}(\pm)}({\bf k})$ is defined as the BCP density profile. 
Here, we can also expand the integrand of $\bar{v}_{xyy}^{\rm BCP}$ under large $\gamma\!=\!v$ as follows:
\begin{eqnarray}
P_{xyy}^{{\rm BCP}(\pm)}({\bf k})
\!\approx\!\pm\!\!\!\lim_{\gamma=v\to\infty}\!\!\!{\rm Re}\!\!\left[\!\frac{k_{x}}{4v(k^{2} \!-\! 1)^{5/2}} \!\right]\!.\label{eq:BCP_v_L}
\end{eqnarray}
This expansion \eqref{eq:BCP_v_L} demonstrates that $\lim_{\gamma=v\to\infty}\bar{v}_{xyy}^{\rm BCP}\!=\!0$,  indicating a complete suppression of the intrinsic response at large non-Hermiticity.
Moreover, we plot the BCP density profile $P_{xyy}^{{\rm BCP}(-)}({\bf k})$ in Eq.~\eqref{eq:BCP_v} in the $k_{x}$--$k_{y}$ plane as shown in Figs.~\ref{fig:BCP_P_t02_g_v_together_M5}(a2)-\ref{fig:BCP_P_t02_g_v_together_M5}(d2). Similar to the BCD density profile, near the exceptional ring at $k^{2}\!=\!R^{2}\!=\!M/\alpha\!=\!1$ where the bands coalesce, the BCP density profile also markedly diverges. The details for the scaling behavior of the BCP density profile near the exceptional ring can be found as follows.
Here, we illustrate that a detectable intrinsic nonlinear Hall response in our model requires a particular scaling law for the BCP density profile. We expand the BCP density profile near the exceptional ring along $k^{2}\!=\!(\sqrt{M/\alpha}\!+\!\delta k)^{2}$ with $\gamma\!=\!v\sqrt{M/\alpha}$ and $\delta k\!\to\!0$, where $\delta k$ denotes a small deviation from the exceptional ring. Along the direction of $k_{y}=0$, we further expand the BCP density profile under $k^{2}\!=\!k_{x}^{2}\!+\!k_{y}^{2}\!=\!(\sqrt{M/\alpha}\!+\!\delta k_{x})^{2}\!+\!(0\!+\!\delta k_{y})^{2}$ to the lowest order of $\delta k_{x}$ and the lowest order of $\delta k_{y}$ as follows:
\begin{eqnarray}
&&P_{xyy}^{{\rm BCP}(\pm)}(\delta k_x,\delta k_y) \nonumber\\
&\!\approx\!&\pm{\rm Re}\!\!\left[\frac{\sqrt{M/\alpha}(v^{2} \!+\! 2M\alpha \!-\! iv\sqrt{M\alpha})}{16\sqrt{2v}\left(v\sqrt{M/\alpha} - 2iM \right)^{5/2}}\!\right]\!\!(\delta k_x)^{-5/2} \nonumber\\
&&\!\pm{\rm Re}\!\!\left[\frac{\sqrt{M/\alpha}(v^{2} \!+\! 2M\alpha \!-\! iv\sqrt{M\alpha})}{4\sqrt{v}\left(v - 2i\sqrt{M\alpha} \right)^{5/2}}\!\right]\!\!(\delta k_y)^{-5}.\label{eq:Pxyy_BCP_dkx}
\end{eqnarray}

Besides, Fig.~\ref{fig:BCP_max_t02_g_v_M_analytic_together} also illustrates that the maximum $\bar{v}_{xyy}^{\rm BCP}$ appears to diverge when $\gamma\!=\!v$ is relatively small but nonnegligible. To understand this, we then consider the above condition of $0\ll\gamma^{2}\!=\!v^{2}\ll1$ and expand Eq.~\eqref{eq:BCP_v} near the exceptional ring along $k\!=\!\sqrt{M/\alpha}\!+\!\delta k\!=\!1\!+\!\delta k$ and $k_{x}\!=\!(1\!+\!\delta k)\cos\theta$ with $M\!=\!\alpha$ and $\delta k\to0$ as follows:
\begin{eqnarray}
P_{xyy}^{{\rm BCP}(\pm)}\!\propto\!\sqrt{\frac{1}{v}}\!=\!\sqrt{\frac{1}{\gamma}}. \label{eq:BCP_v_S3}
\end{eqnarray} Details on the derivation of Eq.~\eqref{eq:BCP_v_S3} can be found in Appendix~\ref{Appendix_11}.
However, this approximation becomes less accurate when $v$ is exceedingly small in the initial $\delta k$ approximation. Numerically, the smallest $\delta k$ is determined by the resolution of the $k$-intervals, which is, in turn, governed by the physical size of the system. For finite lattices, the maximum value of $\bar{v}_{xyy}^{\rm BCP}$ scales as $\sqrt{1/\gamma}$, down to the spatial resolution set by the lattice size.

Interestingly beyond this regime, in the true limit of $\gamma\!=\!v\!\to\!0$, the maximum $\bar{v}_{xyy}^{\rm BCP}$ approaches zero, mirroring the behavior of the BCD.  As such, we can also expand the integrand of Eq.~\eqref{eq:BCP_v} as follows:
\begin{eqnarray}
\lim_{\gamma=v\to0}\!\!P_{xyy}^{{\rm BCP}(\pm)}({\bf k})
\!\approx\!\pm\!\!\lim_{\gamma=v\to0}\!\!{\rm Re}\!\left[\!\frac{k_{x}v^{2}\alpha(3\alpha k^{2} \!-\! M)}{4(M\!-\!\alpha k^{2})^{5}} \!\right]\!,\label{eq:BCP_v0} \nonumber\\
\end{eqnarray} which shows that $\lim_{\gamma=v\to0}\bar{v}_{xyy}^{\rm BCP}\!=\!0$.
Details on the derivation of Eq.~\eqref{eq:BCP_v0} can be found in Appendix~\ref{Appendix_12}.

\section{Discussion}\label{section5}

There are several viable mechanisms to engineer non-Hermiticity in solid-state systems. One common approach is to couple the material to external reservoirs, such as a substrate or a metallic contact, which introduces non-Hermitian effects through energy and particle exchange. In this context, the parameter $\gamma$ naturally arises in the non-Hermitian self-energy. Notable examples include the realization of non-Hermicity in electronic mesoscopic heterojunctions~\cite{geng2023nonreciprocal}, in bilayer graphene systems~\cite{shao2024non}, in altermagnet-ferromagnet junctions~\cite{reja2024emergence}, in topological insulator-ferromagnet junctions~\cite{bergholtz2019non}, and in superconductor-ferromagnet junctions~\cite{cayao2022exceptional}. In addition, artificial platforms such as metamaterials~\cite{ghatak2020observation,fan2022hermitian} and photonic systems~\cite{miri2019exceptional,feng2017non,ozdemir2019parity,zhu2020photonic} offer more direct control over gain and loss, thus providing an accessible and tunable route to explore and manipulate the parameter $\gamma$.

\section{Conclusion}\label{section6}

In this work, we explore how an exceptional ring, characterized by a macroscopic number of defective states, can profoundly impact the nonlinear response, focusing on both extrinsic and intrinsic nonlinear Hall effects. Utilizing an ansatz tilted dissipative Dirac model that hosts an exceptional ring band structure, our findings reveal that the BCD in the extrinsic Hall effect intensifies with increasing non-Hermiticity. In contrast, the intrinsic nonlinear velocity, quantified by the BCP, is significantly suppressed as non-Hermiticity grows, leading to a diminished intrinsic nonlinear response. These behaviors are unique to exceptional rings and unveil novel nonlinear response phenomena arising from the interplay between non-Hermitian defectiveness and extended exceptional band crossings.

\begin{acknowledgements}
We acknowledge helpful discussions with Xiao-Bin Qiang, Hao-Jie Lin, and Rui Chen. F.Q. acknowledges the support from the Jiangsu Specially-Appointed Professor Program in Jiangsu Province and the Doctoral Research Start-Up Fund of Jiangsu University of Science and Technology. C.H.L. acknowledges support from the Singapore Ministry of Education Academic Research Fund Tier-I Grant (WBS No. A-8002656-00-00) and Tier-II Grant (Award No. MOE-T2EP50222-0003)
\end{acknowledgements}

\onecolumngrid
\flushbottom
\appendix

\section{Derivation of Eq.~\eqref{eq:BCP_abc} for the BCP}\label{Appendix_2}

In this Appendix, we present the detailed derivation of Eq.~\eqref{eq:BCP_abc} as follows:
\begin{eqnarray}
\bar{v}_{abc}^{\rm BCP}
&\!=\!&\!\!\sum_{n,n\rq{}}^{\epsilon_{\bf k}^{(n)}\neq \epsilon_{\bf k}^{(n\rq{})}}\!\Gamma_{bc}\!\int\!\!\frac{d^{2}\mathbf{k}}{(2\pi)^{2}}\!\!\left[\frac{v_{a}^{(n)}{\cal G}_{bc}^{(n)}\!-\!v_{b}^{(n)}{\cal G}_{ac}^{(n)}}{\epsilon_{\bf k}^{(n)}\!-\!\epsilon_{\bf k}^{(n\rq{})}}\right]\!\!\frac{\partial f_{0}^{(n)}}{\partial\epsilon_{\bf k}^{(n)}} \nonumber\\
&\!=\!&\!\!\sum_{n,n\rq{}}^{\epsilon_{\bf k}^{(n)}\neq \epsilon_{\bf k}^{(n\rq{})}}\!\Gamma_{bc}\!\int\!\!\frac{d^{2}\mathbf{k}}{(2\pi)^{2}}\!\!\left[\frac{(\partial_{k_{a}}f_{0}^{(n)}){\cal G}_{bc}^{(n)}\!-\!(\partial_{k_{b}}f_{0}^{(n)}){\cal G}_{ac}^{(n)}}{\epsilon_{\bf k}^{(n)}\!-\!\epsilon_{\bf k}^{(n\rq{})}}\right]\!\! \nonumber\\
&\!=\!&\!\!\sum_{n,n\rq{}}^{\epsilon_{\bf k}^{(n)}\neq \epsilon_{\bf k}^{(n\rq{})}}\!\Gamma_{bc}\!\int\frac{dk_b}{2\pi}\!\left[\frac{f_{0}^{(n)}{\cal G}_{bc}^{(n)}}{\epsilon_{\bf k}^{(n)}\!-\!\epsilon_{\bf k}^{(n\rq{})}}\right]\bigg|_{k_a=-\pi}^{k_a=\pi} \!-\! \!\!\sum_{n,n\rq{}}^{\epsilon_{\bf k}^{(n)}\neq \epsilon_{\bf k}^{(n\rq{})}}\!\Gamma_{bc}\!\int\frac{d^{2}{\bf k}}{(2\pi)^2}\!\left[f_{0}^{(n)}\partial_{k_a}\!\!\left(\frac{{\cal G}_{bc}^{(n)}}{\epsilon_{\bf k}^{(n)}\!-\!\epsilon_{\bf k}^{(n\rq{})}}\right)\right]\!\! \nonumber\\
&&\!+\! \sum_{n,n\rq{}}^{\epsilon_{\bf k}^{(n)}\neq \epsilon_{\bf k}^{(n\rq{})}}\!\Gamma_{bc}\!\int\frac{dk_a}{2\pi}\!\left[\frac{-f_{0}^{(n)}{\cal G}_{ac}^{(n)}}{\epsilon_{\bf k}^{(n)}\!-\!\epsilon_{\bf k}^{(n\rq{})}}\right]\bigg|_{k_b=-\pi}^{k_b=\pi} \!-\! \!\!\sum_{n,n\rq{}}^{\epsilon_{\bf k}^{(n)}\neq \epsilon_{\bf k}^{(n\rq{})}}\!\Gamma_{bc}\!\int\frac{d^{2}{\bf k}}{(2\pi)^2}\!\left[-f_{0}^{(n)}\partial_{k_b}\!\!\left(\frac{{\cal G}_{ac}^{(n)}}{\epsilon_{\bf k}^{(n)}\!-\!\epsilon_{\bf k}^{(n\rq{})}}\right)\right]\!\! \nonumber\\
&\!=\!&\sum_{n,n\rq{}}^{\epsilon_{\bf k}^{(n)}\neq \epsilon_{\bf k}^{(n\rq{})}}\!\Gamma_{bc}\!\int\frac{dk_b}{2\pi}\!\left[\frac{f_{0}^{(n)}{\cal G}_{bc}^{(n)}}{\epsilon_{\bf k}^{(n)}\!-\!\epsilon_{\bf k}^{(n\rq{})}}\right]\bigg|_{k_a=-\pi}^{k_a=\pi} 
\!+\! \sum_{n,n\rq{}}^{\epsilon_{\bf k}^{(n)}\neq \epsilon_{\bf k}^{(n\rq{})}}\!\Gamma_{bc}\!\int\frac{dk_a}{2\pi}\!\left[\frac{-f_{0}^{(n)}{\cal G}_{ac}^{(n)}}{\epsilon_{\bf k}^{(n)}\!-\!\epsilon_{\bf k}^{(n\rq{})}}\right]\bigg|_{k_b=-\pi}^{k_b=\pi} \nonumber\\
&&\!-\! \sum_{n,n\rq{}}^{\epsilon_{\bf k}^{(n)}\neq \epsilon_{\bf k}^{(n\rq{})}}\!\Gamma_{bc}\!\int\!\!\frac{d^{2}\mathbf{k}}{(2\pi)^{2}}f_{0}^{(n)}\!\!\left[\!\frac{\partial_{k_a}{\cal G}_{bc}^{(n)}\!-\!\partial_{k_b}{\cal G}_{ac}^{(n)}}{\epsilon_{\bf k}^{(n)}\!-\!\epsilon_{\bf k}^{(n\rq{})}}\!\right] \nonumber\\
&&\!+\! \sum_{n,n\rq{}}^{\epsilon_{\bf k}^{(n)}\neq \epsilon_{\bf k}^{(n\rq{})}}\!\Gamma_{bc}\!\int\!\!\frac{d^{2}\mathbf{k}}{(2\pi)^{2}}\frac{f_{0}^{(n)}\left[{\cal G}_{bc}^{(n)}\!\partial_{k_a}\!\!\left(\epsilon_{\bf k}^{(n)}\!-\!\epsilon_{\bf k}^{(n\rq{})}\right) \!-\! {\cal G}_{ac}^{(n)}\!\partial_{k_b}\!\!\left(\epsilon_{\bf k}^{(n)}\!-\!\epsilon_{\bf k}^{(n\rq{})}\right)\right]}{\left(\epsilon_{\bf k}^{(n)}\!-\!\epsilon_{\bf k}^{(n\rq{})}\right)^{2}}, 
\end{eqnarray} where $\Gamma_{bc}\!=\!2{\cal S}\mathcal{F}_{b}\mathcal{F}_{c}/\hbar$ with the system area being ${\cal S}$, $\mathcal{F}_{b/c}$ is the perturbing force field and  $v_{a}^{(n)}\!=\!\partial\epsilon_{\bf k}^{(n)}/\partial k_{a}$ is the group velocity in the $n$th band~\cite{zhuang2024intrinsic}.

\section{Inversion symmetry}\label{Appendix_3}

The purpose of this Appendix is to demonstrate that the Hamiltonian \eqref{eq:H} in the main text retains inversion symmetry when $t_{0}=0$, and breaks inversion when $t_0\neq 0$. Our Hamiltonian~\eqref{eq:H} under inversion transformation can be expressed as follows:
\begin{eqnarray}
{\cal I}\hat{\cal H}({\bf k}){\cal I}^{-1}
\!=\! t_{0}k_{x}\sigma_{0} \!-\! vk_{y}\sigma_{x} \!-\! \eta vk_{x}\sigma_{y} \!+\! (i\gamma\!+\!M\!-\!\alpha k^{2})\sigma_{z} 
\!=\! \hat{\cal H}(-{\bf k}) \!+\! 2t_{0}k_{x}\sigma_{0},\label{eq:inversion}
\end{eqnarray}
where ${\cal I}\!=\!\sigma_{z}$~\cite{chen2015magnetoinfrared} is the inversion operator, and we use
\begin{eqnarray}
&&(\sigma_{z})(\sigma_{0})(\sigma_{z})^{-1}\!=\!\sigma_{0},\\
&&(\sigma_{z})(\sigma_{x})(\sigma_{z})^{-1}\!=\!-\sigma_{x} ,\\
&&(\sigma_{z})(\sigma_{y})(\sigma_{z})^{-1}\!=\!-\sigma_{y} ,\\
&&(\sigma_{z})(\sigma_{z} )(\sigma_{z})^{-1}\!=\!\sigma_{z}.
\end{eqnarray} According to Eq.~\eqref{eq:inversion}, we find the condition ${\cal I}\hat{\cal H}({\bf k}){\cal I}^{-1}\!=\!\hat{\cal H}(-{\bf k})$ under $t_{0}\!=\!0$, which confirms the inversion symmetry. However, if $t_{0}\neq 0$, then Eq.~\eqref{eq:inversion} indicates that this inversion symmetry is broken.

\section{Time-reversal symmetry is broken}\label{Appendix_4}

This Appendix aims to establish that the Hamiltonian \eqref{eq:H} in the main text does not exhibit time-reversal symmetry, regardless of whether $t_{0}$ is zero or not.

The Hamiltonian~\eqref{eq:H} under time-reversal transformation can be expressed as follows
\begin{eqnarray}
{\cal T}\hat{\cal H}({\bf k}){\cal T}^{-1}
&\!=\!& t_{0}k_{x}\sigma_{0} \!-\! vk_{y}\sigma_{x} \!-\! \eta vk_{x}\sigma_{y} \!-\! (i\gamma\!+\!M\!-\!\alpha k^{2})\sigma_{z} \nonumber\\
&\!=\!&  \hat{\cal H}(-{\bf k}) \!+\! 2t_{0}k_{x}\sigma_{0} \!-\! 2(i\gamma\!+\!M\!-\!\alpha k^{2})\sigma_{z},\label{eq:time_reversal}
\end{eqnarray}
where ${\cal T}\!=\!i\sigma_{y}{\cal K}$~\cite{chen2015magnetoinfrared} is the time-reversal operator with the complex conjugate operator ${\cal K}$. Following this, we have ${\cal K}\hat{\cal H}({\bf k}){\cal K}^{-1}\!=\!\hat{\cal H}^{*}({\bf k})$, with the following relations:
\begin{eqnarray}
&&(i\sigma_{y})(\sigma_{0})(i\sigma_{y})^{-1}\!=\!\sigma_{0},\\
&&(i\sigma_{y})(\sigma_{x})(i\sigma_{y})^{-1}\!=\!-\sigma_{x} ,\\
&&(i\sigma_{y})(K\sigma_{y}K^{-1})(i\sigma_{y})^{-1}\!=\!-\sigma_{y} ,\\
&&(i\sigma_{y})(\sigma_{z} )(i\sigma_{y})^{-1}\!=\!-\sigma_{z}.
\end{eqnarray}
As a result, Eq.~\eqref{eq:time_reversal} demonstrates that the time-reversal symmetry is broken, since ${\cal T}\hat{\cal H}({\bf k}){\cal T}^{-1} \neq \hat{\cal H}(-{\bf k})$ whether $t_{0}$ is zero.

\section{Perturbation near the exceptional ring}\label{Appendix_5}

To investigate the behavior of our model near the exceptional ring, we expand the Hamiltonian \eqref{eq:H} near the exceptional ring. Here, we focus on the expansion along $k^{2}\!=\!(\sqrt{M/\alpha}\!+\!\delta k)^{2}$ with the radius of the ring given by $R=\sqrt{M/\alpha}$ and $\gamma\!=\!v\sqrt{M/\alpha}$. We consider $\delta k\to0$, representing a slight deviation from the ring.

Specifically, along the direction with $k_{y}=0$, the Hamiltonian expansion at  $k^{2}\!=\!k_{x}^{2}\!+\!k_{y}^{2}\!=\!(\sqrt{M/\alpha}\!+\!\delta k_{x})^{2}\!+\!(0\!+\!\delta k_{y})^{2}$ (i.e., near the exceptional ring)  to the lowest order of $\delta k_{x}$ and $\delta k_{y}$ is given by:
\begin{eqnarray}
\hat{\cal H}(\delta k_{x},\delta k_{y}) 
\!\approx\!t_{0}\left(\!\sqrt{\frac{M}{\alpha}}\!+\!\delta k_{x}\!\right)\sigma_{0}\!+\!v(\delta k_{y})\sigma_{x}\!+\!\eta v\left(\!\sqrt{\frac{M}{\alpha}}\!+\!\delta k_{x} \!\right)\sigma_{y} \!+\left[iv\sqrt{\frac{M}{\alpha}}\!-\!2\sqrt{M\alpha}(\delta k_{x})\right]\sigma_{z},\label{eq:H_Taylor}
\end{eqnarray} where $\delta k_{x}\to0$ and $\delta k_{y}\to0$.

Moreover, the corresponding eigenenergies for the model Hamiltonian \eqref{eq:H} near the exceptional ring can be expanded with respect to the lowest order of $\delta k_{x}$ as follows:
\begin{equation}
\epsilon_{\delta k_{x}}^{(\pm)}\!\approx\!t_{0}\left(\!\sqrt{\frac{M}{\alpha}}\!+\!\delta k_{x}\!\right)\!\pm\!\sqrt{2v\left(\! v\sqrt{\frac{M}{\alpha}}\!-\! 2iM \!\right)\delta k_{x}},\label{eq:energy_Taylor}
\end{equation} where we have ignored the higher-order $(\delta k_{y})^{2}$ terms in the square root.

\section{Derivations of Eq.~\eqref{eq:BC} for the form of the Berry curvature}\label{Appendix_6}

In this Appendix, we show the detailed derivations of Eq.~\eqref{eq:BC} as follows~\cite{Berry_connection}:
\begin{eqnarray}
\Omega^{(\pm)}_{z}
&\!=\!&\varepsilon_{zxy}\Omega^{(\pm)}_{xy}\!=\!\varepsilon_{zxy}{\rm Re}\left(\partial_{k_{x}}{\cal A}_{\pm}^{y} \!-\! \partial_{k_{y}}{\cal A}_{\pm}^{x}\right) \nonumber\\
&\!=\!&{\rm Re}\left\{\varepsilon_{zxy}i\left[\!\left\langle \!\frac{\partial\psi_{L}^{(\pm)}}{\partial k_{x}}\!\bigg|\!\frac{\partial\psi_{R}^{(\pm)}}{\partial k_{y}}\!\!\right\rangle \!-\! \left\langle \!\frac{\partial\psi_{L}^{(\pm)}}{\partial k_{y}}\!\bigg|\!\frac{\partial\psi_{R}^{(\pm)}}{\partial k_{x}}\!\!\right\rangle\right]\right\} \nonumber\\
&\!=\!&{\rm Re}\left\{\varepsilon_{zxy}i\sum_{n\rq{}=+,-}\left[\!\left\langle \!\frac{\partial\psi_{L}^{(\pm)}}{\partial k_{x}}\!\left|\!\psi_{R}^{(n\rq{})}\!\right\rangle\!\left\langle\!\psi_{L}^{(n\rq{})}\!\right| \!\frac{\partial\psi_{R}^{(\pm)}}{\partial k_{y}}\!\!\right\rangle \!-\! \left\langle \!\frac{\partial\psi_{L}^{(\pm)}}{\partial k_{y}}\!\left|\!\psi_{R}^{(n\rq{})}\!\right\rangle\!\left\langle\!\psi_{L}^{(n\rq{})}\!\right|\!\frac{\partial\psi_{R}^{(\pm)}}{\partial k_{x}}\!\!\right\rangle\right]\right\} \nonumber\\
&\!=\!&{\rm Re}\left\{\varepsilon_{zxy}i\left[\!\left\langle \!\frac{\partial\psi_{L}^{(\pm)}}{\partial k_{x}}\!\left|\!\psi_{R}^{(\mp)}\!\right\rangle\!\left\langle\!\psi_{L}^{(\mp)}\!\right| \!\frac{\partial\psi_{R}^{(\pm)}}{\partial k_{y}}\!\!\right\rangle \!-\! \left\langle \!\frac{\partial\psi_{L}^{(\pm)}}{\partial k_{y}}\!\left|\!\psi_{R}^{(\mp)}\!\right\rangle\!\left\langle\!\psi_{L}^{(\mp)}\!\right|\!\frac{\partial\psi_{R}^{(\pm)}}{\partial k_{x}}\!\!\right\rangle\right]\right\} \nonumber\\
&\!=\!&{\rm Re}\left\{\varepsilon_{zxy}i\left\{\frac{\!\left\langle \!\psi_{L}^{(\pm)}\!\left|\!\frac{\partial\hat{\cal H}}{\partial k_{x}}\right|\!\psi_{R}^{(\mp)}\!\right\rangle\!\!\left\langle\!\psi_{L}^{(\mp)}\!\left|\!\frac{\partial\!\hat{\cal H}}{\partial k_{y}}\!\right|\!\psi_{R}^{(\pm)}\!\right\rangle}{\left[\epsilon_{\mathbf{k}}^{(\pm)}-\epsilon_{\mathbf{k}}^{(\mp)}\right]^{2}} 
\!-\! \frac{\!\left\langle \!\psi_{L}^{(\pm)}\!\left|\!\frac{\partial\hat{\cal H}}{\partial k_{y}}\right|\!\psi_{R}^{(\mp)}\!\right\rangle\!\!\left\langle\!\psi_{L}^{(\mp)}\!\left|\!\frac{\partial\!\hat{\cal H}}{\partial k_{x}}\!\right|\!\psi_{R}^{(\pm)}\!\right\rangle}{\left[\epsilon_{\mathbf{k}}^{(\pm)}-\epsilon_{\mathbf{k}}^{(\mp)}\right]^{2}} \right\}\right\} \nonumber\\
&\!=\!&\pm{\rm Re}\left\{\frac{\eta v^2 \left(i\gamma + M + \alpha k^2\right)}{2 \left[\left(i\gamma + M - \alpha k^2\right)^2+v^2 k^2\right]^{3/2}} \right\}, \label{eq:BC_A}
\end{eqnarray} where ${\cal A}_{+}^{a}\!=\!i\langle\psi_{L}^{(+)}|\partial_{k_{a}}|\psi_{R}^{(+)}\rangle$ is the $a$ component of the biorthogonal Berry connection~\cite{Berry_connection} ${\cal A}_{+}\!=\!i\langle\psi_{L}^{(+)}|\nabla_{\bf k}|\psi_{R}^{(+)}\rangle$, and we have used $\langle\partial_{k_{a}}\psi_{L}^{(+)}|\psi_{R}^{(-)}\rangle\!=\!\left\langle \!\psi_{L}^{(+)}\!\left|\!\frac{\partial\hat{\cal H}}{\partial k_{a}}\right|\!\psi_{R}^{(-)}\!\right\rangle/\left[\epsilon_{\mathbf{k}}^{(+)}-\epsilon_{\mathbf{k}}^{(-)}\right]$ and $\langle\psi_{L}^{(-)}|\partial_{k_{a}}\psi_{R}^{(+)}\rangle\!=\!\left\langle \!\psi_{L}^{(-)}\!\left|\!\frac{\partial\hat{\cal H}}{\partial k_{a}}\right|\!\psi_{R}^{(+)}\!\right\rangle/\left[\epsilon_{\mathbf{k}}^{(+)}-\epsilon_{\mathbf{k}}^{(-)}\right]$.

Furthermore, we can expand the non-Hermitian Berry curvature \eqref{eq:BC_A} under $k^{2}=(\sqrt{M/\alpha}+\delta k)^{2}$ (i.e., near the exceptional ring) with respect to the lowest order of $\delta k$ as follows:
\begin{eqnarray}
\Omega^{(\pm)}_{z}
\!\approx\!\pm{\rm Re}\left[\!\frac{ i\eta v(\delta k)^{-3/2}}{4\sqrt{2v\left(v\sqrt{\frac{M}{\alpha}}\!-\! 2iM\right)}} \!\right],\label{eq:BC_Taylor}
\end{eqnarray} where we have used $\gamma=v\sqrt{M/\alpha}$ and $\delta k\to0$.

\section{Biorthogonal quantum metric}\label{Appendix_7}

Here, we show the detailed derivations of the biorthogonal quantum metric for any two-band model Hamiltonian as follows:
\begin{eqnarray}
{\cal G}^{(+)}_{ab}&\!\equiv\!&\frac{1}{2}{\rm Re}\left[{\cal Q}^{(+)}_{ab} \!+\! {\cal Q}^{(+)}_{ba} \right] \nonumber\\
&\!=\!&\frac{1}{2}{\rm Re}\left[\langle\partial_{k_a}\psi_{L}^{(+)}|\partial_{k_b}\psi_{R}^{(+)}\rangle \!+\! \langle\partial_{k_b}\psi_{L}^{(+)}|\partial_{k_a}\psi_{R}^{(+)}\rangle \right] 
\!-\! \frac{1}{2}{\rm Re}\left[\! \langle\partial_{k_a}\psi_{L}^{(+)}|\hat{P}^{(+)}|\partial_{k_b}\psi_{R}^{(+)}\rangle \!+\! \langle\partial_{k_b}\psi_{L}^{(+)}|\hat{P}^{(+)}|\partial_{k_a}\psi_{R}^{(+)}\rangle \!\right] \nonumber\\
&\!=\!&\frac{1}{2}{\rm Re}\left[\! \langle\partial_{k_a}\psi_{L}^{(+)}|\hat{P}^{(-)}|\partial_{k_b}\psi_{R}^{(+)}\rangle \!+\! \langle\partial_{k_b}\psi_{L}^{(+)}|\hat{P}^{(-)}|\partial_{k_a}\psi_{R}^{(+)}\rangle \!\right] \nonumber\\
&\!=\!&\frac{1}{2}{\rm Re}\left[\! (\langle\partial_{k_a}\psi_{L}^{(+)}|\hat{P}^{(-)})\hat{P}^{(-)}(\hat{P}^{(-)}|\partial_{k_b}\psi_{R}^{(+)}\rangle) \!+\! (\langle\partial_{k_b}\psi_{L}^{(+)}|\hat{P}^{(-)})\hat{P}^{(-)}(\hat{P}^{(-)}|\partial_{k_a}\psi_{R}^{(+)}\rangle) \!\right] \nonumber\\
&\!=\!&\frac{1}{2}{\rm Re}\left[\! \langle\psi_{L}^{(+)}|\partial_{k_a}\hat{P}^{(+)}|\psi_{R}^{(-)}\rangle\langle\psi_{L}^{(-)}|\partial_{k_b}\hat{P}^{(+)}|\psi_{R}^{(+)}\rangle \!+ \langle\psi_{L}^{(-)}|\partial_{k_a}\hat{P}^{(+)}|\psi_{R}^{(+)}\rangle\langle\psi_{L}^{(+)}|\partial_{k_b}\hat{P}^{(+)}|\psi_{R}^{(-)}\rangle \!\right] \nonumber\\
&\!=\!&\frac{1}{2}{\rm Re}{\rm Tr}\!\left[(\partial_{k_a}\hat{P}^{(+)})(\partial_{k_b}\hat{P}^{(+)})\right]
\!=\!\frac{1}{8}{\rm Re}{\rm Tr}\!\left[(\partial_{k_a}\hat{\bf d}\cdot\boldsymbol{\sigma})(\partial_{k_b}\hat{\bf d}\cdot\boldsymbol{\sigma})\right]\!=\!\frac{1}{4}{\rm Re}[(\partial_{k_a}\hat{\bf d})\cdot(\partial_{k_b}\hat{\bf d})],
\end{eqnarray}
where $\hat{P}^{(\pm)}\!=\!|\psi_{R}^{(\pm)}\rangle\langle\psi_{L}^{(\pm)}|\!=\!\frac{1}{2}\left(\mathbb{I} \pm \hat{\bf d}\cdot\boldsymbol{\sigma} \right)$~\cite{zhuang2023extrinsic} is the biorthogonal projection operator of the ``$\pm$\rq{}\rq{} band, $\hat{\bf d}\!=\!(d_{0}/d,d_{x}/d,d_{y}/d,d_{z}/d)$ with $d\!=\!(d_{x}^{2}\!+\!d_{y}^{2}\!+\!d_{z}^{2})^{1/2}$, $\boldsymbol{\sigma}\!=\!(\sigma_{0},\sigma_{x},\sigma_{y},\sigma_{z})$, $[\sigma_{a},\sigma_{b}]\!=\!2i\varepsilon_{abc}\sigma_{c}$, $\{\sigma_{a},\sigma_{b}\}\!=\!2\delta_{ab}\mathbb{I}$ with the identity matrix $\mathbb{I}$, $a,b\in \{x,y,z\}$, 
$\sum_{n\rq{}=\pm}|\psi_{R}^{(n\rq{})}\rangle\langle\psi_{L}^{(n\rq{})}|\!=\!\mathbb{I}$, we use $\hat{P}^{(-)}|\partial_{k_b}\psi_{R}^{(+)}\rangle\!=\!\partial_{k_b}\hat{P}^{(+)}|\psi_{R}^{(+)}\rangle$~\cite{chen2024quantum}, $\sigma_{a}\sigma_{b}\!=\!i\varepsilon_{abc}\sigma_{c} \!+\! \delta_{ab}\mathbb{I}$ and
\begin{eqnarray}
{\rm Tr}\!\left[(\partial_{k_a}\hat{\bf d}\cdot\boldsymbol{\sigma})(\partial_{k_b}\hat{\bf d}\cdot\boldsymbol{\sigma})\right]
&\!=\!&\frac{1}{d}{\rm Tr}\!\left[(\partial_{k_a}d_{0}\sigma_{0}\!+\!\partial_{k_a}d_{x}\sigma_{x}\!+\!\partial_{k_a}d_{y}\sigma_{y}\!+\!\partial_{k_a}d_{z}\sigma_{z})(\partial_{k_b}d_{0}\sigma_{0}\!+\!\partial_{k_b}d_{x}\sigma_{x}\!+\!\partial_{k_b}d_{y}\sigma_{y}\!+\!\partial_{k_b}d_{z}\sigma_{z})\right] \nonumber\\
&\!=\!&\frac{1}{d}{\rm Tr}[(\partial_{k_a}d_{0})(\partial_{k_b}d_{0})\!+\!(\partial_{k_a}d_{x})(\partial_{k_b}d_{x})\!+\!(\partial_{k_a}d_{y})(\partial_{k_b}d_{y})\!+\!(\partial_{k_a}d_{z})(\partial_{k_b}d_{z})]\mathbb{I} \nonumber\\
&\!=\!&2(\partial_{k_a}\hat{\bf d})\cdot(\partial_{k_b}\hat{\bf d}).
\end{eqnarray} We also have ${\cal G}^{(-)}_{ab}\!=\!{\cal G}^{(+)}_{ab}$.

\section{Scaling behavior of the BCD density profile near the exceptional ring}\label{Appendix_8}

In this Appendix, we will illustrate the scaling behavior of the BCD density profile near the exceptional ring. 
We demonstrate that detecting an extrinsic nonlinear Hall response in our model requires a specific scaling behavior of the BCD density profile. To analyze this, we expand the BCD density profile near the exceptional ring along $k^{2}\!=\!(\sqrt{M/\alpha}+\delta k)^{2}$, with $\gamma\!=\!v\sqrt{M/\alpha}$ and $\delta k\!\to\!0$, where $\delta k$ represents a small deviation from the exceptional ring. Along the $k_{y}\!=\!0$ direction, we further expand the BCD density profile under the condition $k^{2}\!=\!k_{x}^{2}\!+\!k_{y}^{2}\!=\!(\sqrt{M/\alpha}\!+\!\delta k_{x})^{2}\!+\!(0\!+\!\delta k_{y})^{2}$. To the lowest-order terms in $\delta k_{x}$ and $\delta k_{y}$, this expansion takes the following form:
\begin{eqnarray}
P_{yxx}^{(\pm)}(\delta k_{x},\delta k_{y})\!\approx\!\mp\frac{3\eta}{8}{\rm Re}\!\!\left[\!\frac{-v\left(v\sqrt{M/\alpha} \!+\! 2iM \right)}{2(M/\alpha)\left( v^{2} \!+\! 4M\alpha \right)}\!\right]^{1/2}\!\!(\delta k_x)^{-5/2}
\mp\frac{3\eta M}{2\alpha}{\rm Re}\!\!\left[\frac{v}{i2\sqrt{M\alpha}\!-\!v}\!\right]^{1/2}\!\!(\delta k_y)^{-5}.\label{eq:Pyxx_dkx}
\end{eqnarray} 

\section{Scaling behavior of the BCD with $0\ll|\gamma|\!=\!|v|\ll\alpha\!=\!M$ for Eq.~\eqref{eq:BCD_v_S4}}\label{Appendix_9}

We consider the condition of $0\ll|\gamma|\!=\!|v|\ll\alpha\!=\!M$ and expand Eq.~\eqref{eq:BC_kx} near the exceptional ring along $k\!=\!\sqrt{M/\alpha}\!+\!\delta k\!=\!1\!+\!\delta k$ and $k_{x}\!=\!(1\!+\!\delta k)\cos\theta$ with $M\!=\!\alpha$ and $\delta k\to0$ as follows:
\begin{eqnarray}
\frac{\partial\Omega_{z}^{(\pm)}}{\partial k_x} 
&\!=\!& \pm\eta[(1\!+\!\delta k)\cos\theta]v^{2} {\rm Re}\!\left\{\!\!\frac{ \!\alpha v\left[v(1\!+\!\delta k)^{2} \!+\! (i\gamma\!+\!\alpha(1\!-\!(1\!+\!\delta k)^{2}))^{2}/v \right]}{v^{5/2}\left[v(1\!+\!\delta k)^{2} \!+\! (i\gamma\!+\!\alpha(1\!-\!(1\!+\!\delta k)^{2}))^{2}/v \right]^{5/2}} \right.\nonumber\\
&&\left. \!- \frac{3[i\gamma\!+\!\alpha(1\!+\!(1\!+\!\delta k)^{2})]\left[v^{2} \!-\!2\alpha(i\gamma\!+\!\alpha(1\!-\!(1\!+\!\delta k)^{2}))\right]}{2v^{5/2}\left[v(1\!+\!\delta k)^{2} \!+\! (i\gamma\!+\!\alpha(1\!-\!(1\!+\!\delta k)^{2}))^{2}/v \right]^{5/2}} \!\!\right\}\!\nonumber\\
&\!\approx\!&\pm\eta[(1\!+\!\delta k)\cos\theta] {\rm Re}\!\left\{\!\frac{\sqrt{v}\alpha}{[v((1\!+\!\delta k)^{2} \!-\! 1) \!+\! 2i\alpha(1\!-\!(1\!+\!\delta k)^{2}) ]^{3/2}} \right.\nonumber\\
&&\left.\!- \frac{3v^{3/2}[i\gamma\!+\!\alpha(1\!+\!(1\!+\!\delta k)^{2})]}{2[v((1\!+\!\delta k)^{2} \!-\! 1) \!+\! 2i\alpha(1\!-\!(1\!+\!\delta k)^{2}) ]^{5/2}} 
\!-\! \frac{3\alpha v^{3/2}[1 \!-\! 2i\alpha/v \!-\! \alpha^{2}[1 \!-\! (1\!+\!\delta k)^{2}]^{2}/v^{2}]}{[v((1\!+\!\delta k)^{2} \!-\! 1) \!+\! 2i\alpha(1\!-\!(1\!+\!\delta k)^{2}) ]^{5/2}} \!\right\}\! \label{eq:BCD_v_S1} \\
&\!\approx\!&\pm\eta[(1\!+\!\delta k)\cos\theta]
{\rm Re}\!\left\{\!\frac{(v/\alpha)^{1/2}}{[2i(1\!-\!(1\!+\!\delta k)^{2}) ]^{3/2}} \!+\! \frac{6i(v/\alpha)^{1/2}}{[2i(1\!-\!(1\!+\!\delta k)^{2}) ]^{5/2}} \right.\nonumber\\
&&\left.\!- \frac{6(v/\alpha)^{3/2}}{[2i(1\!-\!(1\!+\!\delta k)^{2}) ]^{5/2}}
\!-\! \frac{3i(v/\alpha)^{5/2}}{2[2i(1\!-\!(1\!+\!\delta k)^{2}) ]^{5/2}} \!\right\}\! \label{eq:BCD_v_S2} \\
&\!\approx\!&\pm\eta[(1\!+\!\delta k)\cos\theta]
{\rm Re}\!\left\{\!\frac{(v/\alpha)^{1/2}}{[2i(1\!-\!(1\!+\!\delta k)^{2})]^{3/2}} \!+\! \frac{6i(v/\alpha)^{1/2}}{[2i(1\!-\!(1\!+\!\delta k)^{2}) ]^{5/2}} \!\right\}\! \label{eq:BCD_v_S3} \\
&\!\propto\!&(v/\alpha)^{1/2}\!\propto\!\sqrt{\gamma}.\label{eq:BCD_v_S4_2} 
\end{eqnarray}
In the first line that approximates Eq.~\eqref{eq:BCD_v_S1}, higher-order small terms such as $\alpha^{2}[1\!-\!(1\!+\!\delta k)^{2}]^{2}/v$ in both the numerator and denominator are neglected. Subsequently, higher-order small terms such as $\alpha^{2}[1 \!-\! (1\!+\!\delta k)^{2}]^{2}/v^{2}$ in the numerator and $v((1\!+\!\delta k)^{2} \!-\! 1)$ in the denominator are also neglected in the second approximation \eqref{eq:BCD_v_S2}. Furthermore, higher-order small terms such as $(v/\alpha)^{3/2}$ and $(v/\alpha)^{5/2}$ are neglected in the third approximation \eqref{eq:BCD_v_S3}.

\section{Scaling behavior of the BCD at $\gamma\!=\!v\to 0$ for Eq.~\eqref{eq:BC_kx_v0}}\label{Appendix_10}

The scaling behavior of the BCD at $\gamma\!=\!v\to 0$ can be seen by expanding the term $\partial\Omega_{z}^{(\pm)}/\partial k_{x}$ [Eq.~\eqref{eq:BC_kx}] in the limit of $\gamma\!=\!v\to 0$ as follows:
\begin{eqnarray}
\lim_{\gamma=v\to0}\frac{\partial\Omega_{z}^{(\pm)}}{\partial k_x} 
&\!\approx\!& \!\pm\!\!\!\!\lim_{\gamma=v\to0}\!\!\!\!{\rm Re}\!\left\{\!\!\!\frac{\eta k_{x}v^{2} \!\left[2\alpha(M\!-\!\alpha k^{2})^{2} \!+\! 6\alpha(M\!+\!\alpha k^{2})(M\!-\!\alpha k^{2})\right]}{2(M\!-\!\alpha k^{2})^{5}}\!\!\!\right\}\!\nonumber\\
&\!=\!& \!\pm\lim_{\gamma=v\to0}{\rm Re}\!\left\{\!\!\frac{\eta k_{x}v^{2} \!\left[\alpha(M\!-\!\alpha k^{2}) \!+\! 3\alpha(M\!+\!\alpha k^{2})\right]}{(M\!-\!\alpha k^{2})^{4}}\!\!\right\}\! \nonumber\\
&\!=\!& \!\pm\lim_{\gamma=v\to0}{\rm Re}\!\left[\!\frac{2\alpha k_{x}v^{2}(2M\!+\!\alpha k^{2})}{(M\!-\!\alpha k^{2})^{4}}\!\right] \nonumber\\
&\!=\!& \!\pm\lim_{\gamma\to0}{\rm Re}\!\left[\!\frac{2\alpha k_{x}\gamma^{2}(2\!+\!k^{2})}{M^{3}(1\!-\!k^{2})^{4}}\!\right]\!.\label{eq:BC_kx_v0_2}
\end{eqnarray} 

\section{Scaling behavior of the BCP with $0\ll\gamma^{2}\!=\!v^{2}\ll1$ for Eq.~\eqref{eq:BCP_v_S3}}\label{Appendix_11}

We consider the above condition of $0\ll\gamma^{2}\!=\!v^{2}\ll1$ and expand Eq.~\eqref{eq:BCP_v} near the exceptional ring along $k\!=\!\sqrt{M/\alpha}\!+\!\delta k\!=\!1\!+\!\delta k$ and $k_{x}\!=\!(1\!+\!\delta k)\cos\theta$ with $M\!=\!\alpha$ and $\delta k\to0$ as follows:
\begin{eqnarray}
P_{xyy}^{{\rm BCP}(\pm)}({\bf k})&\!\equiv\!&\!-\!\!\left[\!\frac{\partial_{k_x}{\cal G}_{yy}^{(\pm)}\!-\!\partial_{k_y}{\cal G}_{xy}^{(\pm)}}{\epsilon_{\bf k}^{(\pm)}-\epsilon_{\bf k}^{(\mp)}} \right]
\!+\! \left[\frac{{\cal G}_{yy}^{(\pm)}(v_{x}^{(\pm)}\!-\!v_{x}^{(\mp)})\!-\!{\cal G}_{xy}^{(\pm)}(v_{y}^{(\pm)}\!-\!v_{y}^{(\mp)})}{\left(\epsilon_{\bf k}^{(\pm)}\!-\!\epsilon_{\bf k}^{(\mp)}\right)^{2}} \right] \nonumber\\
&\!=\!&\!\pm{\rm Re}\!\left\{\!\frac{[(1\!+\!\delta k)\cos\theta] v^{2}\left[v^{2} \!+\! 3\alpha^{2}(1\!+\!\delta k)^{2} \!-\! \alpha\left(M \!+\! i\gamma\right) \right]}{4v^{5/2}[v(1\!+\!\delta k)^{2} \!+\! [i\gamma\!+\!\alpha(1\!-\!(1\!+\!\delta k)^{2})]^{2}/v]^{5/2}} \!\right\}\! \nonumber\\
&\!\approx\!&\!\pm{\rm Re}\!\left\{\!\!\frac{[(1\!+\!\delta k)\cos\theta]\alpha^{2}[3(1\!+\!\delta k)^{2}\!-\!1]}{4\sqrt{v}[v((1\!+\!\delta k)^{2} \!-\! 1) \!+\! 2i\alpha(1\!-\!(1\!+\!\delta k)^{2}) ]^{5/2}} \!\!\right\}\! \label{eq:BCP_v_S1} \\
&\!\approx\!&\!\pm{\rm Re}\!\left\{\!\frac{[(1\!+\!\delta k)\cos\theta]\alpha^{2}[3(1\!+\!\delta k)^{2}\!-\!1]}{4\sqrt{v}[2i\alpha(1\!-\!(1\!+\!\delta k)^{2}) ]^{5/2}} \!\right\}\! \label{eq:BCP_v_S2}\\
&\!\propto\!&\sqrt{\frac{1}{v}}\!=\!\sqrt{\frac{1}{\gamma}}. \label{eq:BCP_v_S3_2}
\end{eqnarray}
In the first line that approximates Eq.~\eqref{eq:BCP_v_S1}, higher-order small terms such as $v^{4}$ and $iv^{2}\gamma$ in the numerator and $\alpha^{2}[1\!-\!(1\!+\!\delta k)^{2}]^{2}/v$ in the denominator are neglected. Subsequently, the term $v((1\!+\!\delta k)^{2} \!-\! 1)$ in the denominator is also neglected in the second approximation \eqref{eq:BCP_v_S2}. Therefore, we obtain the derived analytical scaling behavior $\sqrt{1/\gamma}$ described by Eq.~\eqref{eq:BCP_v_S2}, which aligns with our numerical results shown in Fig.~\ref{fig:BCP_max_t02_g_v_M_analytic_together}.

\section{Scaling behavior of the BCP at $\gamma\!=\!v\to 0$ for Eq.~\eqref{eq:BCP_v0}}\label{Appendix_12}

In the limit of $\gamma\!=\!v\!\to\!0$, we can expand the integrand of Eq.~\eqref{eq:BCP_v} as follows:
\begin{eqnarray}
\lim_{\gamma=v\to0}P_{xyy}^{{\rm BCP}(\pm)}({\bf k})&\!=\!&\!-\!\!\!\lim_{\gamma=v\to0}\!\left[\!\frac{\partial_{k_x}{\cal G}_{yy}^{(\pm)}\!-\!\partial_{k_y}{\cal G}_{xy}^{(\pm)}}{\epsilon_{\bf k}^{(\pm)}-\epsilon_{\bf k}^{(\mp)}} \right] 
\!+\! \lim_{\gamma=v\to0}\!\left[\frac{{\cal G}_{yy}^{(\pm)}(v_{x}^{(\pm)}\!-\!v_{x}^{(\mp)})\!-\!{\cal G}_{xy}^{(\pm)}(v_{y}^{(\pm)}\!-\!v_{y}^{(\mp)})}{\left(\epsilon_{\bf k}^{(\pm)}\!-\!\epsilon_{\bf k}^{(\mp)}\right)^{2}} \right] \nonumber\\
&\!\approx\!&\pm\!\!\lim_{\gamma=v\to0}\!\!{\rm Re}\!\left[\!\frac{k_{x}v^{2}\alpha(3\alpha k^{2} \!-\! M)}{4(M\!-\!\alpha k^{2})^{5}} \!\right]\!.\label{eq:BCP_v0_2} 
\end{eqnarray}

\section{Expectation value of an observable in a non-Hermitian system}\label{Appendix_13}

While using right eigenvectors on both sides may be more physical when considering a specific state, employing both left and right eigenvectors is more appropriate when the state occupancy is determined by energy. In this case, projecting onto the occupied bands requires a properly defined projector, which must be constructed using the left and right biorthogonal basis.

The reason why the Berry curvature is commonly calculated using left and right eigenvectors in non-Hermitian systems stems from the role of the occupied energy bands and the trace operation, which involves summing over a complete set of states when deriving the expectation value of an observable $\hat{\Omega}$ in a non-Hermitian system. This biorthogonal formulation ensures consistency with physical observables and properly captures the geometric properties of non-Hermitian bands.

A quantum system can be described by the time-independent, unperturbed non-Hermitian Hamiltonian $\hat{H}_{0}$. This means that an expectation value of a physical quantity $\hat{\Omega}$ can be evaluated as
\begin{eqnarray}
\langle\hat{\Omega}\rangle \!=\! \frac{\text{Tr}(\hat{\rho}_{0}^{}\hat{\Omega})}{Z_{0}} \!=\! \frac{1}{Z_{0}}\sum_{n,n'}\langle\psi_{n'}^{L}|\psi_{n}^{R}\rangle\langle\psi_{n}^{L}|\hat{\Omega}|\psi_{n'}^{R}\rangle e^{-\beta E_{n}} \!=\! \frac{1}{Z_{0}}\sum_{n}\langle\psi_{n}^{L}|\hat{\Omega}|\psi_{n}^{R}\rangle e^{-\beta E_{n}}, 
\end{eqnarray}
where $|\psi_{n}^{R}\rangle$ ($|\psi_{n}^{L}\rangle$) is the right (left) eigenvector of the occupied energy band $E_{n}$ in $\hat{\cal H}_{0}$ with $\hat{\cal H}_{0}|\psi_{n}^{R}\rangle\!=\!E_{n}|\psi_{n}^{R}\rangle$ and $\hat{\cal H}_{0}^{\dagger}|\psi_{n}^{L}\rangle\!=\!E_{n}^{*}|\psi_{n}^{L}\rangle$, the summation $\sum_{n}$ in the last term is taken over all occupied bands, $\beta\!=\!1/(k_{B}T)$ with the system temperature $T$ and the Boltzmann constant $k_{B}$, $\hat{\rho}_{0}^{}$ is the density operator, $Z_{0}\!=\!\text{Tr}(\hat{\rho}_{0}^{})$ is the partition function, and
\begin{eqnarray}
\hat{\rho}_{0}^{} &\!=\!& e^{-\beta \hat{H}_{0}} \!=\! e^{-\beta \sum_{n}\hat{\cal H}_{0}|\psi_{n}^{R}\rangle\langle \psi_{n}^{L}|} 
\!=\! e^{-\beta \sum_{n}E_{n}|\psi_{n}^{R}\rangle\langle \psi_{n}^{L}|} \nonumber\\
&\!=\!& \sum_{n',n''}|\psi_{n'}^{R}\rangle\langle \psi_{n'}^{L}|e^{-\beta\sum_{n}E_{n}|\psi_{n}^{R}\rangle\langle \psi_{n}^{L}|} |\psi_{n''}^{R}\rangle\langle \psi_{n''}^{L}| \!=\! \sum_{n}|\psi_{n}^{R}\rangle\langle \psi_{n}^{L}| e^{-\beta E_{n}}.
\end{eqnarray}
Here, we write the density operator in terms of a complete set of eigenstates $\{|\psi_{n}^{R/L}\rangle\}$ of the unperturbed non-Hermitian Hamiltonian $\hat{H}_{0}$ with the eigenenergies $\{E_{n}\}$ and $\hat{H}_{0} \!=\! \sum_{n}\hat{\cal H}_{0}|\psi_{n}^{R}\rangle\langle \psi_{n}^{L}|$. The trace operation, denoted as ${\rm Tr}$, involves summing over a complete set of states:
\begin{eqnarray}
{\rm Tr}(\cdots)\!=\!\sum_{n}\langle\psi_{n}^{L}|\cdots|\psi_{n}^{R}\rangle,
\end{eqnarray} where only the biorthogonal basis gives the resolution of the identity~\cite{brody2013biorthogonal}.

\twocolumngrid
\bibliography{references_nonlinear2}
\end{document}